\documentclass[12pt]{article}

\usepackage{newtxtext,newtxmath}
\usepackage{graphicx}

\usepackage[letterpaper,margin=1in]{geometry}

\renewenvironment{abstract}
	{\quotation}
	{\endquotation}

\date{}

\makeatletter
\renewcommand{\fnum@figure}{\textbf{Figure \thefigure}}
\renewcommand{\fnum@table}{\textbf{Table \thetable}}
\makeatother

\usepackage{scicite}

\usepackage{url}

\def\scititle{%The Evolution of a Multiphase Galactic Wind and its Impact on Lyman Continuum Escape}%Tracking the Time Evolution of a Multiphase Galactic Wind and its Connection to Lyman Continuum Escape}
	Supernovae Driven Winds Impede Lyman Continuum Escape from Dwarf Galaxies in First 10 Myr}
\title{\bfseries \boldmath \scititle}

\author{
	Cody Carr$^{1,2\ast}$,
    Renyue Cen$^{1,2\ast}$,
	Stephan McCandliss$^{3\ast}$,
    Jack Ford$^{3}$,\and
    Alberto Saldana-Lopez$^{4}$,
    Claudia Scarlata$^5$,
    Mason S. Huberty$^5$,
    Anne Jaskot$^{6}$,\and
    Sophia Flury$^{7}$,
    M. S. Oey$^{8}$,
    Ricardo O. Amor\'{i}n$^{9}$,
    Sanchayeeta Borthakur$^{10}$,\and
    Matthew Hayes$^{4}$,
    Timothy Heckman$^{3}$,
    Zhiyuan Ji$^{11}$,
    Lena Komarova$^{8}$,\and
    Alexandra Le Reste$^5$,
    Floriane Leclercq$^{12}$,
    Rui Marques-Chaves$^{13}$,
    Leo Michel-Dansac$^{14}$,\and
    Göran Östlin$^{4}$,
    Swara Ravindranath$^{15,16}$,
    Michael~J.~Rutkowski$^{17}$,
    Daniel Schaerer$^{10}$,\and
    Trinh Thuan$^{18}$,
    Eros Vanzella$^{19}$,
    Bingjie Wang$^{20,21,22}$,
    Xinfeng Xu$^{23}$
\and
	\small$^{1}$Center for Cosmology and Computational Astrophysics, Institute for Advanced Study in Physics,\\ \small Hangzhou 310058, China.\and
	\small$^{2}$Institute of Astronomy, School of Physics, Zhejiang University, Hangzhou 310058, China.\and
    \small$^{3}$Johns Hopkins University, Department of Physics \& Astronomy, Center for Astrophysical Sciences,\\ \small Baltimore 21218, United States.\and
    \small$^{4}$Department of Astronomy, Oskar Klein Centre, Stockholm University, Stockholm 106 91\and
    \small$^{5}$Minnesota Institute for Astrophysics, University of Minnesota, Minneapolis 55455, United States.\and
    \small$^{6}$Williams College, Department of Physics and Astronomy, Williamstown 01267, United States.\and
    \small$^{7}$University of Edinburgh, Royal Observatory, Institute for Astronomy\\ \small Edinburgh EH9 3HJ, United Kingdom,\and
    \small$^{8}$ Astronomy Department, University of Michigan, Ann Arbor 48109, United States\and
    \small$^{9}$ Instituto de Astrof\'{i}sica de Andaluc\'{i}a (CSIC), Apartado 3004, Granada 18080, Spain\and
    \small$^{10}$ School of Earth Space Exploration, Arizona State University, Tempe 85287, United States \and
    \small$^{11}$Steward Observatory, University of Arizona, Tucson 85721, United States\and
    \small$^{12}$ Centre de Recherche Astrophysique de Lyon, Univ Lyon, Saint-Genis-Laval F-69230, France
    \and
    \small$^{13}$ Observatoire de Genève, Université de Genève, Versoix 1290, Switzerland\and
    \small$^{14}$ CNRS, Aix Marseille Univ, Marseille 13388, France \and
    \small$^{15}$ Astrophysics Science Division, NASA Goddard Space Flight Center, Greenbelt 20771, United States \and
    \small$^{16}$ Center for Research and Exploration in Space Science and Technology II, Department of Physics, \\ \small Catholic University of America, Washington DC 20064, United States \and
    \small$^{17}$ Department of Physics \& Astronomy, Minnesota State University-Mankato, Mankato 56001, United States \and
    \small$^{18}$Astronomy Department, University of Virginia, Charlottesville 22904, United States\and
    \small$^{19}$ INAF – OAS, Osservatorio di Astrofisica e Scienza dello Spazio di Bologna, Bologna 40129, Italy\and
    \small$^{20}$ Department of Astronomy \& Astrophysics, The Pennsylvania State University,\\ \small University Park 16802, United States\and
    \small$^{21}$ Institute for Computational \& Data Sciences, The Pennsylvania State University,\\ \small University Park 16802, United States\and
    \small$^{22}$ Institute for Gravitation and the Cosmos, The Pennsylvania State University,\\ \small University Park 16802, United States\and
    \small$^{23}$Department of Physics and Astronomy, Northwestern University, Evanston 60208, United States\and
	\small$^\ast$Corresponding author. Email: codycarr24@gmail.com, renyuecen@zju.edu.cn, stephan@pha.jhu.edu \and
}

\begin{document} 

% Insert the title and author list
\maketitle

% Abstract, in bold
% There are strict length limits, and not all formats have abstracts.
% Consult the journal instructions to authors for details.
% Do not cite any references in the abstract.
\begin{abstract} \bfseries \boldmath
% Start with one or two sentences of background

% Note the abstract is preferred to have < 150 words, but our draft can have a max of 250

Observations suggest that UV-bright, compact star-forming galaxies produce enough ionizing (Lyman continuum; LyC) photons to reionize the Universe. Yet, the efficiency of LyC escape and the roles of radiation, stellar winds, and supernovae remain uncertain. Using medium-resolution spectra of six nearly identical local star-forming galaxies, we directly trace, for the first time, the evolution of a multiphase wind through individual spectral lines alongside measurements of the LyC escape fraction. We find that LyC escape peaks early, during a period dominated by intense radiation and stellar winds but lacking a fast galactic wind. As the starbursts age, supernovae drive and accelerate the wind, progressively suppressing LyC escape. These results highlight the need for cosmological simulations to incorporate early feedback as a key driver of reionization.

%The abundance of UV-bright, compact star-forming galaxies in the early Universe suggests that high-energy photons from massive stars alone could have reionized the intergalactic medium. Yet, the efficiency with which ionizing (Lyman continuum; LyC) photons escape from these galaxies remains poorly constrained. Uncertainty persists over the timing and impact of stellar feedback, including radiation, stellar winds, and supernovae. Using medium-resolution spectra of six nearly identical local star-forming galaxies, we directly trace, for the first time, the evolution of a multiphase wind through individual spectral lines alongside measurements of the LyC escape fraction. We find that LyC escape peaks early, during a period dominated by intense radiation and strong stellar winds but lacking a fast galactic wind.  As the starbursts age, supernovae continuously expel mass from the ISM, accelerating the galactic wind and progressively suppressing LyC escape. These results highlight the need for cosmological simulations to incorporate early feedback as a key driver of reionization.
%Cosmological simulations have long predicted that LyC escape peaks after the onset of supernova-driven winds, which are thought to carve low-density channels through the interstellar medium. Yet recent multiwavelength observations of local galaxies suggest that an earlier period, dominated by radiation feedback, may be more critical. \textbf{}
\end{abstract}

\section{Introduction}

\noindent 
Reionization marks a critical epoch in cosmic history when the first galaxies ionized the majority of the neutral hydrogen that once filled intergalactic space. Yet, the details of how reionization evolved and what the dominant sources of reionization are remain uncertain.  Early observations at low redshift with the Hubble Space Telescope (HST) \cite{Shapley2006,Izotov2016_four,Izotov2016_one,Steidel2018,Fletcher2019,Flury2022_data}, now supported by detections at high redshift with the James Webb Space Telescope (JWST) \cite{Atek2024,Tacchella2023}, point to a population of compact, UV-bright, star-forming galaxies as leading candidates.  Their observable far-ultraviolet (FUV) spectra at high redshifts indicate that these galaxies alone produce enough ionizing (Lyman continuum; LyC, $<912$ \AA) photons to complete reionization by its endpoint—roughly one billion years after the Big Bang \cite{Lin2024,Munoz2024,Pahl2025}. However, it remains unclear what fraction of these photons actually escape the neutral gas and dust—the two primary sinks of LyC radiation—within and surrounding the galaxies to ionize intergalactic space.

Current evidence indicates that stellar feedback—supernovae (SNe), stellar winds, and radiation—is required to create low-density channels that facilitate LyC escape. In the framework commonly adopted by cosmological simulations, LyC escape is expected to peak after the onset of SN-driven feedback \cite{Kimm2014,Ma2020,Rosdahl2022,Choustikov2024}.  SNe can launch powerful galactic winds that accelerate gas, dust, and metals beyond the galaxy’s escape velocity, clearing pathways for LyC photons to escape, typically after the most massive and prodigious LyC-producing stars have died.  In contrast, multiwavelength studies of local LyC emitters suggest that escape may peak early, while massive stars are still present, before the development of large-scale SN-driven winds \cite{Jaskot2017,Jaskot2019,Flury2022_diagnostics,Flury2025_ISM,Jecmen2023,Bait2024,Carr2025_LyC}. In this scenario, LyC photons leak through a porous interstellar medium (ISM), potentially aided by diffuse, radiation-driven outflows \cite{Kakiichi2021,Menon2025,Komarova2021,Komarova2025}. A third possibility invokes a combination of radiation and mechanical feedback, with both young and older stars contributing to substantial LyC escape \cite{Flury2025_ISM}.

Establishing the LyC escape sequence—and the role of stellar feedback in enabling it—is essential for understanding how galaxies reionized the Universe. Yet the timing and impact of galactic winds remain uncertain, with outcomes likely governed by their multiphase structure, dynamics, and dominant driving mechanism. Winds have long been thought to facilitate LyC escape in compact starbursts \cite{Heckman2001,Heckman2011,Alexandroff2015}, and several observational studies report correlations between $f_{\rm esc}^{\rm LyC}$ and wind speed diagnostics \cite{Amorin2024,Li2025,Komarova2021,Komarova2025}. By contrast, other studies suggest that winds can suppress LyC escape by enhancing absorption in the neutral phase \cite{Jaskot2019,Carr2021,Carr2025_LyC}. Many of these conclusions rely on indirect wind diagnostics and model-dependent interpretations, often producing contradictory results from the same data [cf. \cite{Carr2025_LyC,Li2025}; see also \cite{Jaskot2025} for a recent review]. Here, we address these issues in compact star-forming galaxies by directly tracking the LyC escape sequence—comparing similar systems at different evolutionary stages—and capturing the emergence of galactic winds through spectral lines arising in different ionization states.

%The origin of this neutral, LyC-blocking component is itself unclear: theoretical studies indicate that it may either be entrained directly from the ISM into multiphase outflows or condense later out of the hot wind fluid filling the CGM \cite{Fielding2022,Decataldo2024,Carr2025_LyC}. Additional open questions include whether different temperature phases are dynamically coupled or decoupled \cite{Carr2025_LyC}, and whether winds accelerate or decelerate over time \cite{Li2025,Prusinski2025}. 

The absorption of LyC photons by the neutral hydrogen that permeated the intergalactic medium before the end of reionization makes direct observations of LyC escape from the first galaxies impossible. Astronomers therefore study low-redshift analogs of high-redshift galaxies to infer LyC escape in the early universe. In this study, we present new \textit{HST}/COS observations of six compact, highly star-forming galaxies at redshift $z \sim 0.3$. These targets, drawn from the Low-$z$ Lyman Continuum Survey (LzLCS; \cite{Flury2022_data}), were selected to span a wide range in LyC escape fraction ($f_{\rm esc}^{\rm LyC}=0$–20\%), while occupying a narrow range in metallicity ($8.2 \leq 12+\log{\rm O/H} \leq 8.5$), star-formation rate ($1.3 \leq \log{\rm SFR}\ [{\rm M}\odot\ {\rm yr}^{-1}] \leq 1.6$), and stellar mass ($9.13 \leq \log{M\star}\ [{\rm M}_\odot] \leq 9.75$). Aside from these criteria and observational feasibility constraints (see Table~\ref{tab:observations}), the galaxies were selected at random. Detailed properties of each target are listed in Table~\ref{tab:galaxy_props}.

This parameter selection controls for the key physical forces that shape starburst-driven outflows—specifically, the outward forces from supernovae and radiation pressure, and the inward forces from gravity and ram pressure. We review a parameterization of these forces in the relevant terms in the Supplementary Text. As we show, variation in $f_{\rm esc}^{\rm LyC}$ is primarily governed by the development of the neutral ($\lesssim 10^4$ K) phase of the galactic winds, which evolves on $\sim$10 Myr timescales \cite{Kimm2014,Ma2020,Hayes2023,Carr2025_FIRE2}. Therefore, this data set enables a direct comparison of similar galactic systems at different stages of their starburst evolution, offering an unprecedented view into how $f_{\rm esc}^{\rm LyC}$ evolves over the lifetime of a starburst {\it in tandem} with the development of a galactic wind.  Despite this novelty, we note that our conclusions are limited by our small sample size and advocate for deeper, higher resolution spectra of additional LyC emitters.

\section{Results}

We provide stellar population age estimates for each galaxy in Table~\ref{tab:light_fractions}, derived from spectral energy distribution (SED) modeling. Specifically, we present UV light fractions \cite{Flury2025_ISM}—the fraction of total UV luminosity contributed by stellar populations in three age bins: young ($<3$ Myr), intermediate (3–5 Myr), and old (5–10 Myr) and light-weighted ages. This analysis was originally performed at lower spectral resolution using LzLCS data \cite{Saldana-Lopez2022}. Here, we re-fit the SEDs at the higher resolution of our dataset and find significant changes in $f_{\rm esc}^{\rm LyC}$ with an average relative error of $\sim 45\%$. With the exception of galaxies J115855+312559 and J091703+315221 swapping positions, the rank order in $f_{\rm esc}^{\rm LyC}$ remains consistent with the lower-resolution results, as does the general classification of strong ($\geq 20\%$), weak ($5$–$20\%$), and non-leakers ($<5\%$). While these differences do not alter our main conclusions, a larger sample is required to fully understand the impact of spectral resolution on the derived quantities. Details of our modeling procedure are provided in \cite{methods}.

Stellar age estimates also change, though their uncertainties remain substantial. For this reason, we focus on broad trends rather than precise values, and consider additional age diagnostics in the Supplementary Text. In general, the strongest LyC leakers—J103344+635317, J091703+315221, and J115855+312559—tend to host younger stellar populations ($<4$ Myr), with J103344+635317 being the youngest. In contrast, the weakest leakers—J105331+523753, J144010+461937, and J154050+572442—have older populations ($>4$ Myr), with J154050+572442 being the oldest. To mitigate uncertainties in individual age bins, we rank the galaxies by the light fraction from stars older than 3 Myr, which are capable of producing SNe. This ranking yields a sequence that decreases monotonically with $f_{\rm esc}^{\rm LyC}$.

\begin{figure} % Do not use \begin{figure*}
	\centering
	\includegraphics[width=\textwidth]{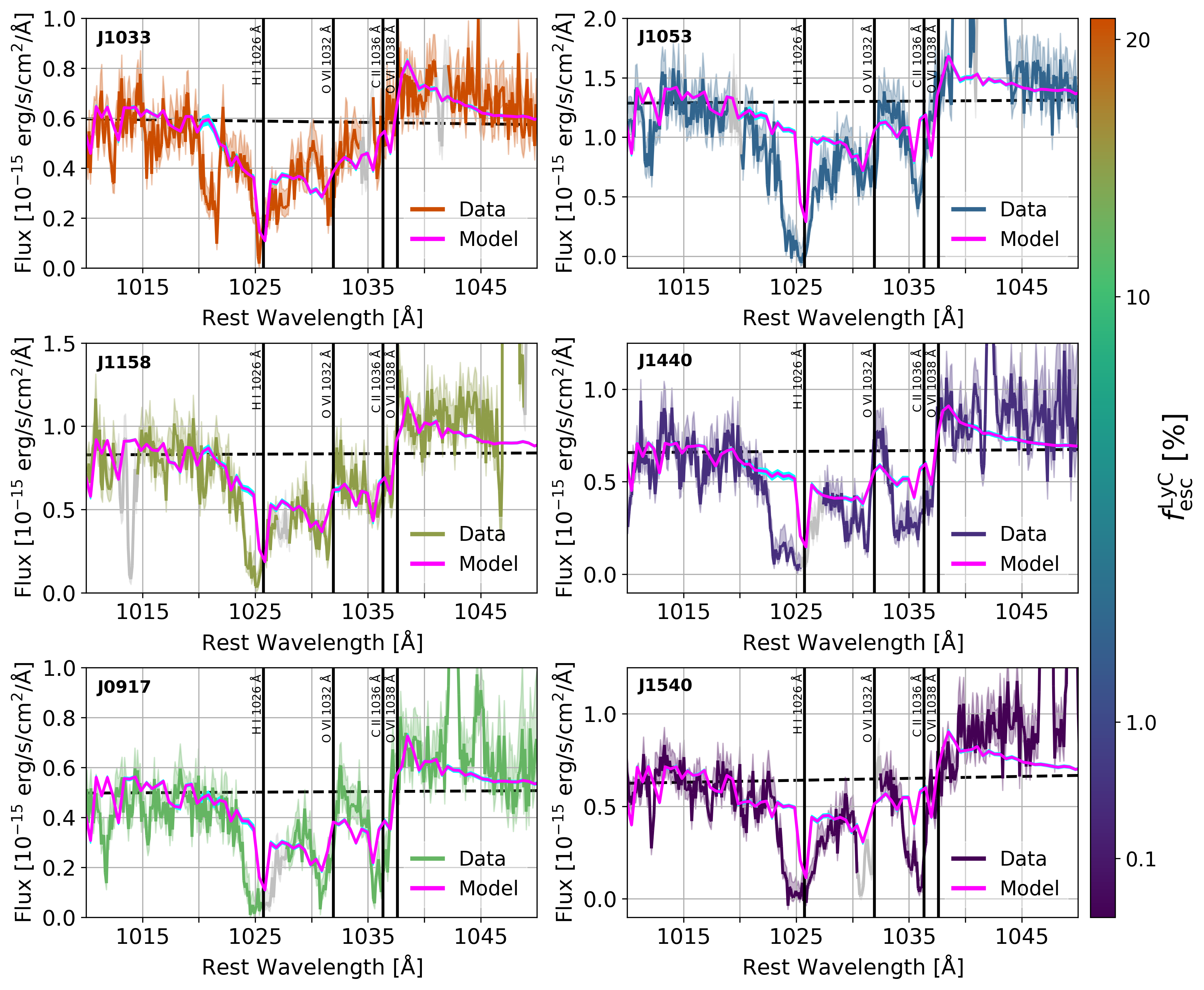} % for an image file named example_figure.*
	% Pick an appriopriate width for the size of the image

	% Captions go below figures
	\caption{\textbf{Time evolution of stellar winds alongside the Lyman continuum escape fraction ($\boldsymbol{f_{\mathrm{esc}}^{\mathrm{LyC}}}$) in compact, highly star-forming galaxies observed in O\,VI~1032\,\AA, 1038\,\AA\ absorption lines.}  Galaxies are arranged in order of decreasing $f_{\rm esc}^{\rm LyC}$ from top to bottom, left to right and are colored according to their individual $f_{\rm esc}^{\rm LyC}$ values, as indicated by the color bar.  Silver regions denote contamination by Milky Way absorption.  The best fitting SED, revealing stellar components, is shown in magenta. Galaxies with higher $f_{\rm esc}^{\rm LyC}$ (e.g., J103344 + 635317, orange) exhibit deep, large-scale ($\sim$ 20 \AA) absorption features, likely corresponding to the fast stellar winds of massive O-type stars.  In contrast, galaxies with lower $f_{\rm esc}^{\rm LyC}$ (e.g., J154050 + 572442, purple), show shallow or absent, large-scale absorption features and enhanced small-scale ($\sim$ 3 \AA) galactic wind absorption features, that fail to be captured by the stellar SED model.  This trend likely reflects the deaths of the most massive O stars and the emergence of SN-driven winds as the starbursts age.}
	\label{fig:stellar_winds} % give each figure a logical label name
\end{figure}

We show spectra and their best-fitting SED models from 1010–1050 \AA\ in Figure~\ref{fig:stellar_winds}, covering the H I 1026 \AA\ (Ly$\beta$) line and O\,VI 1032 \AA, 1038 \AA\ doublet [see also \cite{Izotov2018,Chisholm2022,Flury2025_ISM}]. Our inferred time sequence flows from top to bottom, left to right, with decreasing $f_{\rm esc}^{\rm LyC}$. The model SEDs reveal two key stellar features: (1) a large-scale ($\sim25$ \AA) P Cygni profile, consisting of absorption blueward and emission red-ward of the O VI 1032 \AA\ line, and (2) a small-scale ($\sim 3$ \AA) absorption feature centered near each line. The large-scale features are associated with young ($<3$ Myr), massive O-type stars [see \cite{Chisholm2022}, Figure 1].  The broad absorption feature occurs in the fast ($\sim3000$ km/s, \cite{Morton1979}) winds of these stars, where O VI formation is generally attributed to Auger ionization induced by X-rays from shocks \cite{Hillier2020}. We searched for evidence of Wolf–Rayet stars in the SDSS optical spectra (e.g. He II 4686 \AA), but found no significant broadening typically associated with their fast winds [e.g.
\cite{Rivera-Thorsen2024}]. The small-scale features are attributed to absorption in stellar photospheres. As $f_{\rm esc}^{\rm LyC}$ decreases, the large-scale P Cygni profiles flatten, and the small-scale features no longer account for the full observed absorption around each line, which we attribute to the onset of SN-driven galactic winds. This evolution fits naturally into our proposed time sequence: as starbursts age, massive stars die—either exploding as SNe or collapsing directly into black holes \cite{OConnor2011,Sukhbold2016}—leading to a decline of fast stellar winds and the emergence of fast SN-driven galactic winds in the oldest starbursts, such as J154050+572442.

%The inferred time evolution of the starbursts implies substantial changes in the dominant form of stellar feedback over time. In the youngest systems, feedback is expected to be dominated by radiation and potentially fast stellar winds, preceding the onset of supernovae. As the starbursts age, supernova explosions greatly enhance the energy and momentum budget \cite{Jecmen2023}, launching massive, large-scale galactic winds.  In turn, this evolving feedback plays a pivotal role in shaping the neutral gas and dust distributions within the galaxies and their immediate surroundings, thereby establishing the critical link between the age of the starburst and $f_{\rm esc}^{\rm LyC}$ [see \cite{Jaskot2019,Hayes2023_LyA,Carr2025_LyC,Flury2025_ISM} for more discussion].  We can explore this connection directly by examining the spectral lines arising within the different phases of the outflows in our sample, where outflowing gas reveals itself through blue-shifted absorption features.

\begin{figure} % Do not use \begin{figure*}
	\centering
	\includegraphics[width=\textwidth]{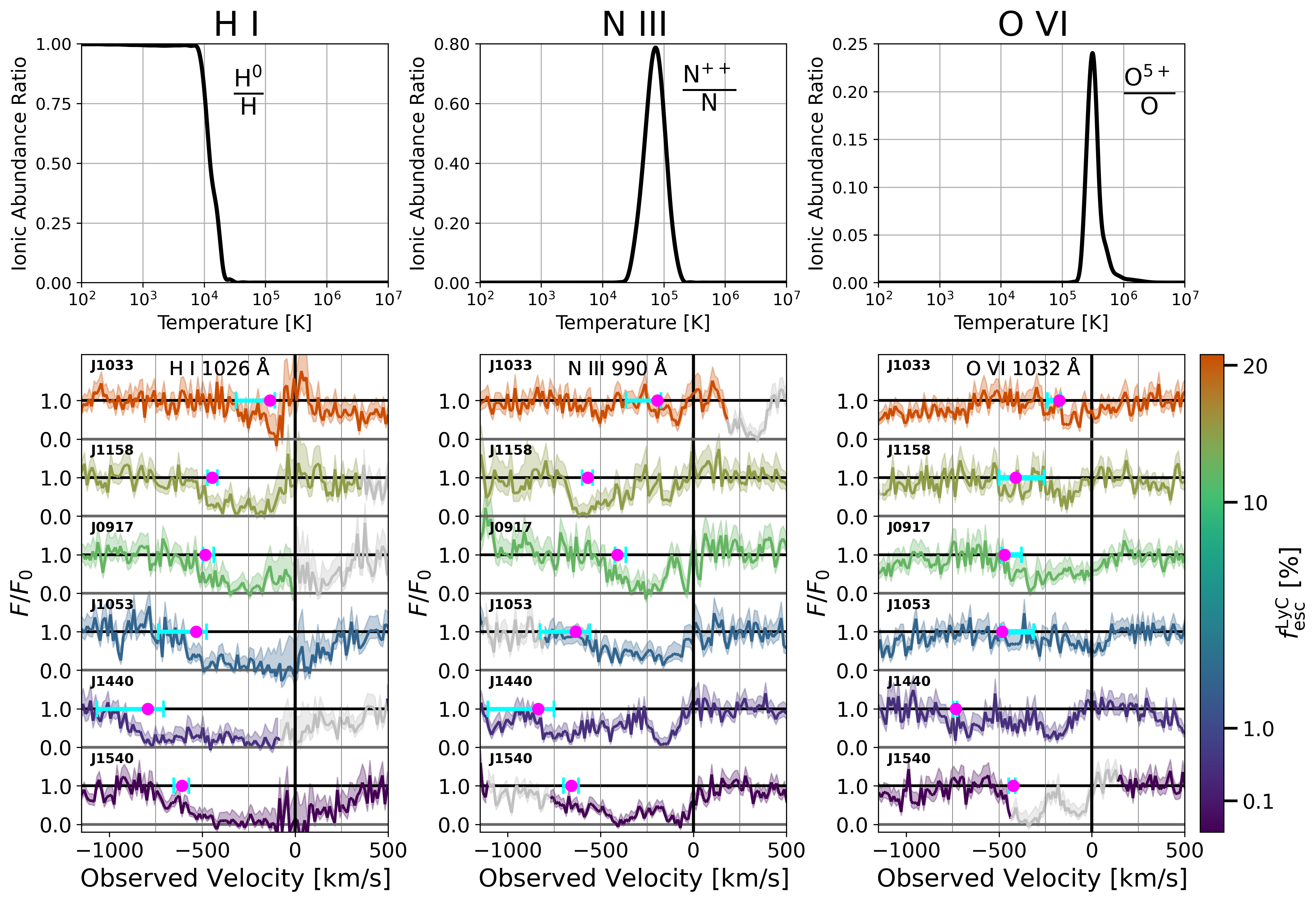} % for an image file named example_figure.*
	% Pick an appriopriate width for the size of the image

	% Captions go below figures
	\caption{\textbf{Time evolution of a multiphase galactic wind and the LyC escape fraction ($\boldsymbol{f_{\rm esc}^{\rm LyC}}$) observed in the spectra of compact, highly star-forming galaxies.}  Each column corresponds to a different ionic line, H I (left), N III (middle), and O VI (right).  The top row shows ionic abundances calculated at a metallicity of $0.4 Z_{\odot}$, redshift $z = 0.3$, hydrogen density $n_{\rm HI} = 1\ \mathrm{cm^{-3}}$, and ionization equilibrium in the presence of intergalactic and extragalactic ionizing radiation, as well as cosmic rays \cite{Ploeckinger2020}.  Together, the three lines trace the wind ionization structure from $\sim 10^2-10^6$ K.  Galaxies are ordered from top to bottom with decreasing $f_{\rm esc}^{\rm LyC}$ and colored according to their $f_{\rm esc}^{\rm LyC}$ values, as indicated by the color bar on the right. Silver regions denote contamination from Milky Way absorption. All spectra are normalized by the best-fitting SED model.  The left, middle, and right bottom panels show the H I 1026 \AA, N III 990 \AA, and O VI 1032 \AA\ lines, respectively.  $v_{90}$ values are shown in magenta with errors in cyan.  In general, both wind speed (width) and mass (depth) increase as $f_{\rm esc}^{\rm LyC}$ declines in all lines.  Except for J115855 + 312559 (N III 990\AA, gold) and J154050 + 572442 (O VI 1032\AA, purple), the ionized outflows closely follow their cooler counterparts, suggesting a multiphase outflow. These results imply that $f_{\rm esc}^{\rm LyC}$ peaks before the onset of SNe and declines following the subsequent mass loading of a SN-driven wind.}
	\label{fig:cool_warm_phases} % give each figure a logical label name
\end{figure}

In the left panel of Figure~\ref{fig:cool_warm_phases}, we compare spectral cutouts, normalized by the best fitting SED model, containing the H I 1026 \AA\ absorption line for each galaxy.  Our inferred time sequence flows from top to bottom with decreasing $f_{\rm esc}^{\rm LyC}$. The temperature regimes probed by the different spectral lines used in this study are shown in the top panels of Figure~\ref{fig:cool_warm_phases}. These values are adopted from \cite{Ploeckinger2020} and computed at $z = 0.3$, metallicity $0.4Z_{\odot}$, and hydrogen density $n_{\rm HI} = 1.0\ \mathrm{cm^{-3}}$ \cite{Carr2025_LyC}, assuming ionization equilibrium in the presence of intergalactic and extragalactic radiation fields as well as cosmic rays. They are meant only to guide the reader’s intuition for the temperature range probed here; deriving precise values for each galaxy would require detailed radiative transfer modeling beyond the scope of this work [e.g., \cite{Gray2019_winds}]. Equivalent widths (EW) and $v_{90}$ values—the velocity at which 90\% of the EW is contained when integrating blue-ward of line center—are listed in Table~\ref{tab:line_data} for all lines and various correlations with H I 1026 \AA\ are explored in Figure~\ref{fig:correlations}.

J103344+635317, our strongest leaker, exhibits the weakest outflow in both mass (depth) and velocity (width) in H I. The presence of a P Cygni profile indicates scattering in a low-density, spherical wind \cite{Carr2018}. In contrast, the galaxies with the lowest $f_{\mathrm{esc}}^{\mathrm{LyC}}$ values—J105331+523753, J144010+461937, and J154050+572442—show significantly faster outflows. Although these galaxies lack P Cygni profiles, they likely still host spherical winds; at high densities, fluorescent scattering via Ly$\alpha$ is expected to suppress such features \cite{Scarlata2015}. Moreover, most galaxies hosting galactic winds in LzLCS exhibit large H I covering fractions ($\sim 1$), and are well modeled as spherical winds \cite{Li2025}. The appearance of redshifted absorption in the weaker leakers suggests continuous mass loading in the winds and an increase in density over time. Indeed, in accelerating winds, this feature can arise from non-resonant photon absorption, enhanced in high-density regions near the galaxy \cite{Carr2023}.

We examine higher-ionization states of the winds using the N III $\lambda990$ and O VI $\lambda1032$ lines, shown in the middle and right panels of Figure~\ref{fig:cool_warm_phases}, respectively. Together, these lines trace gas over a temperature range of roughly $10^4$–$10^6$ K. With the exceptions of J115855+312559 (N III) and J154050+572442 (O VI), both transitions show higher outflow velocities at lower $f_{\rm esc}^{\rm LyC}$ and $v_{90}$ values comparable to the H I wind, consistent with multiphase outflows. The absorption dip in J115855+312559 is not seen in other lines and likely reflects contamination. In contrast, the absence of high-velocity O VI absorption in J154050+572442, one of our oldest starbursts, may indicate that the gas has cooled substantially, dropping below our detection threshold. Although the equivalent widths are generally weaker, the shapes of the N III and O VI line profiles closely resemble those of the H I wind in the four strongest leakers. However, in the two weakest leakers, J154050+572442 and J144010+461937, we detect multiple components or line splitting, indicating a piling-up of material—possibly marking the formation of superbubbles associated with SN-driven winds or bursty star formation.   

These results disfavor the presence of fast ionized winds in J103344 + 635317, consistent with \cite{Jaskot2017,Jaskot2019}, who found that Green Pea galaxies with the lowest optical depths also exhibited the weakest ionized and neutral wind velocities.  However, our analysis does not probe the hot ($\gtrsim 10^7$ K) wind phase, which requires X-ray observations [e.g., \cite{Strickland1997}]. We have X-ray data only for J103344 + 635317, recently observed with XMM-Newton by \cite{Ji2025}. The object was not detected, with a 3-sigma upper limit on the 0.5–8 keV X-ray luminosity of $10^{40.8}\ \mathrm{erg\ s^{-1}}$, suggesting a lack of high-mass X-ray binaries consistent with a young stellar population ($<5$ Myr). See Supplementary Text for further discussion. 

\section{Discussion and Conclusions}

Together, our findings suggest that $f_{\rm esc}^{\rm LyC}$ peaks early in the lifecycle of a starburst and gradually declines as a large-scale, SN-driven outflow develops \cite{Jaskot2019,Flury2022_diagnostics,Carr2021,Carr2025_LyC}. This outflow is multiphase and accelerated as it emerges from the ISM. Continuous mass loading causes these winds to gain in neutral mass and dust over a roughly 10 Myr time period. We further explore the spatial extent of the winds through surface brightness profiles in Figure S3 in the Supplementary Text.

\begin{figure} % Do NOT use \begin{figure*}
	\centering
	\includegraphics[width=\textwidth]{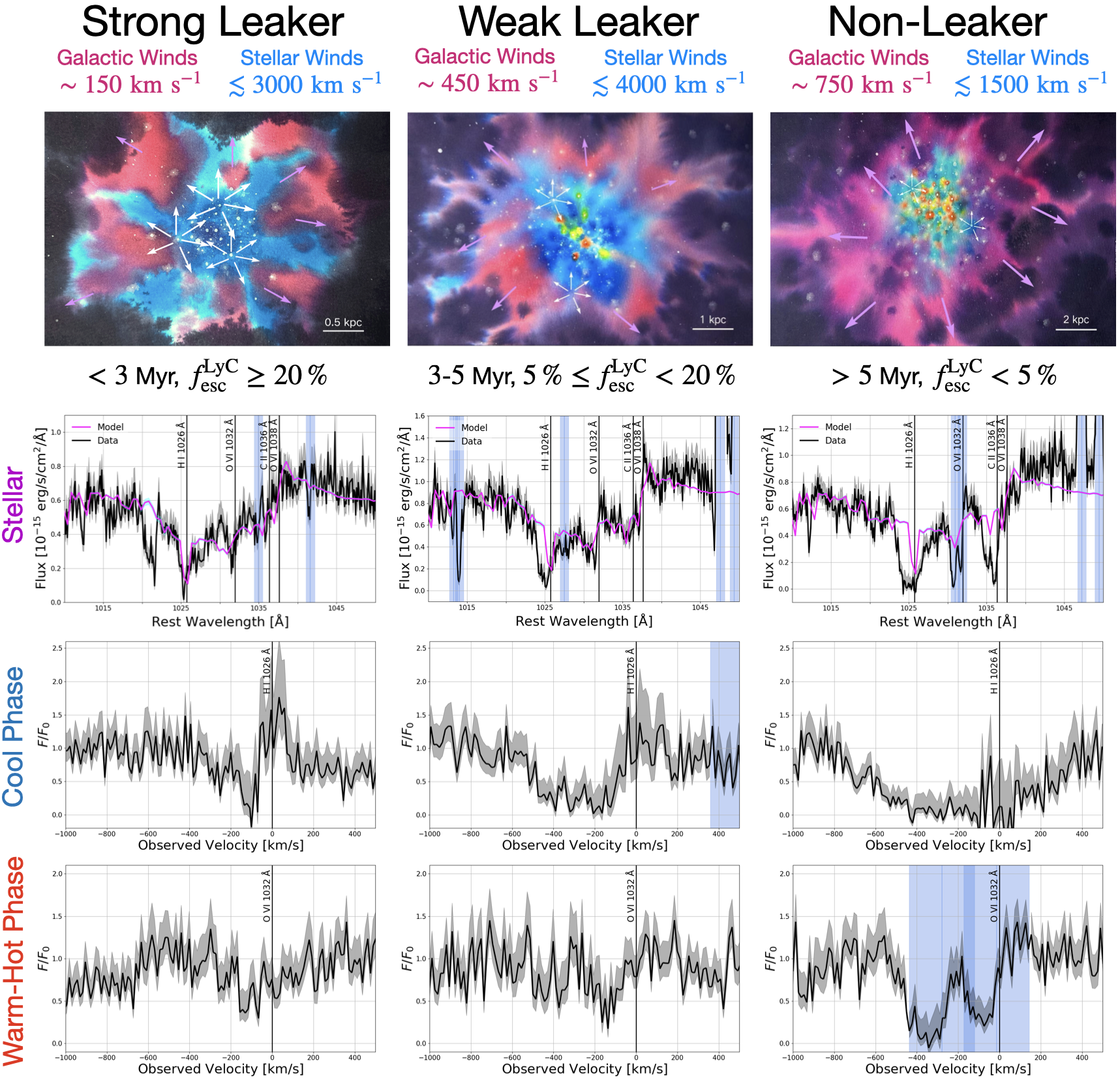} % for an image file named example_figure.*
	% Pick an appropriate width - in print, figures are usually one or two columns wide, which can
	% be approximated by 0.3\textwidth or 0.6\textwidth respectively. Use appropriate label sizes.

	% Captions go below figures
	\caption{\textbf{An artist’s rendering of the LyC escape sequence alongside the evolution of a multiphase wind in compact, star-forming galaxies, as inferred from spectral lines tracing $\mathbf{10^2\text{–}10^6}$ K gas.}
    Example observations are shown, with light blue regions marking Milky Way contamination (see Figures \ref{fig:stellar_winds} and \ref{fig:cool_warm_phases} for details). Red clouds represent the multiphase wind, UV-dominated regions in blue; lighter shades indicate lower densities.  Radiation and fast stellar winds drive a slow wind early, without depositing significant dust or neutral gas.  Gas clumping and intense ionizing feedback promote high $f_{\rm esc}^{\rm LyC}$ during this period.  After the onset of SNe, mass and dust are expelled, lowering $f_{\rm esc}^{\rm LyC}$ on average despite the creation of low-density channels.  At later times, SNe drive large-scale winds that block LyC escape, possibly enhanced by condensation from the warm gas.  See also \cite{Jaskot2019,Hayes2023_LyA,Bait2024,Carr2025_LyC,Flury2022_diagnostics,Flury2025_ISM}.} 
	\label{fig:summary} % give each figure a logical label name
\end{figure}

This scenario supports a volume-filling wind in which radiation and stellar winds—possibly aided by a few early SNe from the most massive stars (see Supplementary Text)—expel the first material from the ISM \cite{Komarova2025}.  Early feedback and clumping open low-density pathways with little dust or neutral gas, creating optimal conditions for LyC escape \cite{Jecmen2023, Flury2022_data}.  As more SNe occur, particularly in lower-mass stellar populations, they drive additional mass out of the ISM and produce dust.  These SNe likely destroy dense clouds and carve lower-density channels \cite{Flury2025_ISM}, but the accompanying mass loading, dust creation, and increased lateral volume filling of the wind reduce $f_{\rm esc}^{\rm LyC}$ on average \cite{Jaskot2019,Carr2021,Carr2025_LyC}. Over time, the rising column density of gas and dust suppresses LyC escape entirely, with the neutral phase likely enhanced by cooling in warmer gas \cite{Voit2021}. This decline is compounded by the death of the most massive, LyC-producing stars in the oldest starbursts. A schematic of this evolutionary sequence is shown in Figure~\ref{fig:summary}. %We include the fast ($\sim4000~\mathrm{km,s^{-1}}$) winds of Wolf–Rayet stars in the sequence for reference, even though we do not detect them [cf. \cite{Rivera-Thorsen2024}]. Nevertheless, the energy and momentum budget driving the winds is expected to be dominated by SNe after 
%5 -- 8 Myr \cite{Jecmen2023}.
% 3 Myr.   

Our results are broadly consistent with \cite{Carr2025_LyC}, who used radiative transfer modeling of Mg II lines to map neutral and partially ionized winds in LzLCS galaxies. While we detect multiphase winds, they proposed that the neutral component may be kinematically decoupled from the warmer phases in strong leakers such as J103344+635317. This scenario was introduced to reconcile the lack of Mg II absorption with the observed correlation between $f_{\rm esc}^{\rm LyC}$ and the [O~III] broad-line wings—possible tracers of galactic winds—reported by \cite{Amorin2024,Komarova2025}. However, our findings appear to contradict these trends. Our results also differ from those of \cite{Li2025}, who performed a similar analysis as \cite{Carr2025_LyC} and found that the strongest leakers in LzLCS exhibit the fastest Mg II winds. We compare our velocity measurements to those of \cite{Komarova2025,Li2025} for galaxies with overlapping coverage in Figure~\ref{fig:vel_comp}, with further discussion in the Supplementary Text.

%However, our results appear to contradict these trends. We explore this more thoroughly in the Supplementary Text and consider the possibility of early, diffuse, radiation-driven superwinds that remain undetected in absorption \cite{Komarova2021,Komarova2025}.

One possible explanation may be that we are not observing the full LyC escape sequence. After the most massive stars die, LyC production from binary products—such as blue stragglers and stripped stars—can dominate for 10–50 Myr \cite{Secunda2020}. During this period, galactic winds may begin to clear the ISM and enable LyC escape rather than suppress it, particularly if mass loading is reduced. This could explain some of the fast-wind, low-LyC galaxies observed in the correlation by \cite{Komarova2025}, some of which also lack very young ($<3$ Myr) stars. Feedback from earlier starbursts could also pre-clear the ISM and CGM, facilitating LyC escape in subsequent episodes \cite{Martin2024}. Finally, we note that two of our strongest leakers, J091703 + 315221 and J115855 + 312559, host moderately fast ($\sim 500\ \rm km\ s^{-1}$) galactic winds. Considered in isolation [e.g., \cite{Rivera-Thorsen2017,Vanzella2022}], such systems might suggest that winds facilitate LyC escape. Our sequence, however, indicates that the neutral phase of the winds in these galaxies is not yet developed enough to fully block LyC escape. To fully understand these relationships, similarly high-resolution spectra of the remaining LzLCS galaxies will be required to capture the full diversity of the sample.

These observations contrast with current cosmological hydrodynamical simulations, which typically predict LyC escape to peak following the onset of SNe \cite{Kimm2014,Ma2020,Rosdahl2022,Choustikov2024}. The discrepancy may arise from a failure to resolve H II regions or missing subgrid physics, including optimal radiation hydrodynamics \cite{Deng2024} and stellar winds \cite{Marinacci2019}. Capturing the effect of SNe-driven winds may also require higher resolution in the CGM, as simulations with forced high mass resolution typically show higher H I covering fractions at moderate column densities ($\log N < 18$ [cm${}^{-2}$]), still capable of modulating $f_{\rm esc}^{\rm LyC}$ \cite{Peeples2019}. Converging in $f_{\rm esc}^{\rm LyC}$ in cosmological simulations remains notoriously challenging \cite{Ma2020,Smith2022,Rosdahl2022}.

Idealized simulations of giant molecular clouds (GMCs) suggest that the short dynamical times of compact galaxies naturally lead to rapid star formation and high $f_{\rm esc}^{\rm LyC}$ early during periods of intense radiation feedback, consistent with our observations \cite{Menon2025}. In these environments, strong radiation from young stellar populations can ionize and clear low-column-density channels that naturally occur in turbulent media \cite{Kakiichi2021}. Stellar cluster–scale physics is also likely important: catastrophic cooling and pressure confinement of weak winds can suppress mechanical feedback while enhancing clumping \cite{Silich2018,Jaskot2019,Gray2019_cooling,Danehkar2021,Danehkar2022,Oey2023}, potentially explaining the combination of high escape fractions and low outflow rates observed in strong leakers such as J103344 + 635317. 

For more discussion on the agreement between simulations and local observations, see the review by \cite{Jaskot2025}.

These results have important implications for cosmic reionization. Radiation feedback is expected to be stronger in early compact galaxies, many of which may be super-Eddington \cite{Ferrara2024}. Their low metallicities—only a few percent solar—are expected to lead to more efficient and prolonged radiation-dominated phases, as massive stars collapse directly into black holes, delaying the onset of SN-driven winds \cite{Jecmen2023,Renzini2023,Carr2025_LyC,Flury2025_ISM}. At the same time, stellar winds are expected to be weaker at low metallicity \cite{Vink2022}, suggesting that radiation feedback dominates \cite{Jecmen2023,Komarova2025}. Whether compact, star-forming galaxies were sufficient to reionize the Universe remains an open question, but identifying the key physical processes that regulate LyC escape is a critical first step.

The deep spectroscopic dataset analyzed in this study was only obtainable with HST, underscoring the importance of similar observations as the telescope approaches the end of its operational lifetime. Future observatories, such as the Habitable Worlds Observatory, will enable routine low-redshift observations at high spectral resolution, while 30-meter-class ground-based telescopes will extend these studies to even higher redshifts. Together, this and other high-resolution spectroscopic studies will provide critical benchmarks for these next-generation instruments.

\newpage

%%%%%%%%%%%%%%%% MAIN TEXT FIGURES %%%%%%%%%%%%%%%

%%%%%%%%%%%%%%%% MAIN TEXT TABLES %%%%%%%%%%%%%%%

\begin{table}
    \centering
    \caption{\textbf{Physical properties of the six local star-forming galaxies.}
    The columns show the object name, UV half-light radius $R_{\rm UV}$, stellar mass $M_\star$, log of the H$\beta$-derived star formation rate (SFR), gas-phase metallicity $12 + \log({\rm O/H})$, and Lyman continuum escape fraction $f_{\rm esc}^{\rm LyC}$. Quoted uncertainties and upper limits reflect the 1$\sigma$ errors.  All values were obtained from \cite{Flury2022_data,Flury2022_data_erratum}, except for $f_{\rm esc}^{\rm LyC}$, which was measured in this study.}
    \label{tab:galaxy_props}
    
    \begin{tabular}{lccccc}
        \hline
        Object & $R_{\rm UV}$ & $\log M_\star$ & $\log{\mathrm{SFR}}$ & $12 + \log({\rm O/H})$ & $f_{\rm esc}^{\rm LyC}$ \\
        & (kpc) & $(\log_{10} M_\odot)$ & $\log{(M_\odot\,\mathrm{yr}^{-1})}$ & (dex) & (\%) \\
        \hline
        J103344+635317 & $0.544 \pm 0.029$ & $9.13^{+0.44}_{-0.43}$ & $1.379 \pm 0.025$ & $8.242 \pm 0.039$ & $21.0^{2.2}_{2.0}$ \\
        J115855+312559 & $0.546 \pm 0.029$ & $9.75^{+0.52}_{-0.36}$ & $1.328 \pm 0.021$ & $8.386 \pm 0.031$ & $14.7^{3.7}_{1.5}$ \\
        J091703+315221 & $0.407 \pm 0.029$ & $9.31^{+0.44}_{-0.43}$ & $1.293 \pm 0.020$ & $8.459 \pm 0.035$ & $12.0^{2.1}_{1.5}$ \\
        J105331+523753 & $0.618 \pm 0.030$ & $9.29^{+0.44}_{-0.43}$ & $1.438 \pm 0.020$ & $8.251 \pm 0.032$ & $2.2^{2.4}_{0.7}$ \\
        J144010+461937 & $0.641 \pm 0.029$ & $9.55^{+0.44}_{-0.43}$ & $1.552 \pm 0.020$ & $8.206 \pm 0.033$ & $0.4^{0.5}_{0.1}$ \\
        J154050+572442 & $0.946 \pm 0.031$ & $9.62^{+0.44}_{-0.43}$ & $1.397 \pm 0.025$ & $8.396 \pm 0.043$ & $<0.4$ \\
        \hline
    \end{tabular}
\end{table}

\begin{table}
    \centering
    \caption{\textbf{Fractional light contributions from different aged stellar populations and the mean or light weighted age.} The columns show the object name, the fractional contribution to the total UV light from stars in a given age range in Myr, and the light weighted age.}
    \label{tab:light_fractions}
    
    \begin{tabular}{lccccc}
        \hline
        Object & 0--3 Myr & 3--5 Myr & 5--10 Myr & 4--10 Myr&Mean Age\\
               &      &      &   &  & (Myr) \\
        \hline
        J103344+635317 & 0.72$\pm$0.20 & 0.17$\pm$0.10 & 0.11$\pm$0.06 & 0.28$\pm$0.12&2.71$\pm$0.75\\
        J115855+312559 & 0.59$\pm$0.17  & 0.32$\pm$0.18 & 0.10$\pm$0.07  &0.41$\pm$0.19& 3.39$\pm$1.11\\
        J091703+315221 & 0.47$\pm$0.20 & 0.53$\pm$0.20   & 0.00$\pm$0.04  &0.53$\pm$0.21& 3.70$\pm$1.13\\
        J105331+523753 & 0.43$\pm$0.12 & 0.00$\pm$0.06  & 0.57$\pm$0.10 & 0.57$\pm$0.12& 5.38$\pm$0.94 \\
        J144010+461937 & 0.34$\pm$0.18  & 0.60$\pm$0.16   & 0.07$\pm$0.14  & 0.66$\pm$0.21& 3.97$\pm$1.50\\
        J154050+572442 & 0.33$\pm$0.05  & 0.13$\pm$0.08 & 0.55$\pm$0.08 & 0.67$\pm$0.11& 5.43$\pm$0.83\\
        \hline
    \end{tabular}
\end{table}

%%%%%%%%%%%%%%%% REFERENCES %%%%%%%%%%%%%%%

\clearpage % Clear all remaining figures and tables then start a new page

% The list of references goes after the main text and before the acknowledgements
% When preparing an initial submission, we recommend you use BibTeX, like this:
%
\bibliography{radiation_escape_pathways} % for a file named science_template.bib
\bibliographystyle{sciencemag}

% After the paper has completed peer review and been revised ready for acceptance,
% you should comment out the lines above and copy-paste the contents of your .bbl
% file here instead. This will help ensure that our conversion software works correctly.
% Remember to re-run BibTeX first - check the timestamp!
%
% Example of the first three entries copy-pasted from science_template.bbl:
%
%\begin{thebibliography}{1}
%
%\bibitem{example}
%A.~N. {Author}, An example reference. \emph{Journal of Improbable Research}
%  \textbf{1}, 67 (2020).
%
%\bibitem{example2}
%F.~M. {Surname}, S.~{Author}, A second example. \emph{Interesting Research
%  Letters} \textbf{32}, 897 (2019).
%
%\bibitem{example_preprint}
%P.~{One}, P.~{Two}, P.~{Three}, {An unpublished preprint}. \emph{preprint}
%  (2021), arXiv:2101.12345.
%
%\end{thebibliography}

%%%%%%%%%%%%%%%% ACKNOWLEDGEMENTS %%%%%%%%%%%%%%%

\section*{Acknowledgments}
We thank the entire LzLCS team for the many insightful discussions and support over the years.  Lastly, we thank artist Siyi Wang for bringing our interpretations of the galaxies in Figure~\ref{fig:summary} to life through watercolor painting, and for gracefully incorporating our many, sometimes conflicting, adjustments with great aesthetic sensitivity.  The analysis presented in this article was in part carried out on the SilkRiver supercomputer of Zhejiang University.  OpenAI ChatGPT (version 4.0) was used to assist with language and phrasing of the manuscript. 
%Here you can thank helpful colleagues who did not meet the journal's authorship criteria, or
%provide other acknowledgements that don't fit the (compulsory) subheadings below.
%Formatting requirements for each of these sections differ between the \textit{Science}-family
%journals; consult the instructions to authors on the journal website for full details.
\paragraph*{Funding:}
C.~C. is supported by NSFC grant W2433001 and the NSFC
Talent-Introduction Program.  R.~C. acknowledges in part financial support from the start-up funding of Zhejiang University and Zhejiang provincial top level research support program. S.~M. acknowledges support from HST-GO-17443.002-A to JHU.  S.~B. acknowledges support from NSF grant 2408050.  F.~L. acknowledges funding from the European Union's Horizon 2020 research and innovation program under the Marie Skodowska-Curie grant agreement No. C3UBES - 101107619.   
%List the grants, fellowships etc. that funded the research;
%use initials to specify which author(s) were supported by each source.
%Include grant numbers when appropriate or required by the funding agency.
%For example: F.~A. was funded by the Generous Science Agency grant~2372.
\paragraph*{Author contributions:}
Following the CRediT (Contribution Roles Taxonomy) system, the main roles of the authors were:

C.~C. Conceptualization, data curation, formal analysis, funding acquisition, investigation, methodology, software, visualization, writing -- original draft

R.~C. Conceptualization, funding acquisition, project administration, resources, methodology, supervision, writing -- original draft

S.~M. Conceptualization, data curation, formal analysis, funding acquisition, investigation, methodology, software, supervision, validation, visualization, writing -- original draft

A.~S. Formal analysis, methodology, software, writing – review \& editing

J.~F. Formal analysis, methodology, software, validation, writing – review \& editing

C.~S. writing -- review \& editing 

M.~H. writing -- review \& editing 

A.~J. writing -- review \& editing 

S.~F. writing -- review \& editing 

M.~S.~O. writing -- review \& editing

R.~O.~A. writing -- review \& editing 

S.~B. writing -- review \& editing

M.~H. writing -- review \& editing

T.~H. writing -- review \& editing 

Z.~J. writing -- review \& editing 

L.~K. writing -- review \& editing 

A.~L. writing -- review \& editing 

F.~L. writing -- review \& editing 

R.~M.—C. writing -- review \& editing 

L.~M.—D. writing -- review \& editing 

G.~O. writing -- review \& editing 

S.~R. writing -- review \& editing 

M.~J.~R. writing -- review \& editing 

D.~S. writing -- review \& editing 

T.~T. writing -- review \& editing   

E.~V. writing -- review \& editing

B.~W. writing -- review \& editing 

X.~X. writing -- review \& editing 

\paragraph*{Competing interests:}

There are no competing interests to declare.

%Disclose any potential conflicts of interest for all authors, such as patent applications,
%additional affiliations, consultancies, financial relationships etc.
%See the journal editorial policies web page for types of competing interest that must be declared.
%If there are no competing interests, state:
%``There are no competing interests to declare.''
\paragraph*{Data and materials availability:}

All HST/COS spectral data used in this study are publicly available through the Mikulski Archive for Space Telescopes (MAST) at the Space Telescope Science Institute (STScI). Data for the six galaxy targets analyzed in this paper correspond to the following HST program IDs: 12928, 14201, 17153, 17443. The data can be accessed via the MAST portal at \url{https://mast.stsci.edu} using these program identifiers. Additional photometric measurements from the LaCOS survey (program 17069) are included in the same archive. XMM-Newton data for J103344+635317 can be found at the European Space Archive at \url{https://nxsa.esac.esa.int/nxsa-web/} under program ID: 0934050101.  The radiative transport code used in our analysis is publicly available at \url{https://semi-analytic-line-transfer-salt.readthedocs.io} or \url{https://github.com/CodyCarr/SALT}. There are no restrictions on data or code reuse.

\subsection*{Supplementary materials}
Materials and Methods\\
Supplementary Text\\
Figs. S1 \\
Tables S1, S2 \\
References \textit{(69-\arabic{enumiv})}\\ % automatically fills out the last reference number
% (filling out the other numbers automatically is possible but fiddly and liable to break)
%Movie S1\\
%Data S1

%%%%%%%%%%%%%%%% END OF MAIN TEXT %%%%%%%%%%%%%%%

\newpage

%%%%%%%%%%%%%%%% START OF SUPPLEMENT %%%%%%%%%%%%%%%

% Figures, tables, equations and pages in the supplement are numbered S1, S2 etc.
\renewcommand{\thefigure}{S\arabic{figure}}
\renewcommand{\thetable}{S\arabic{table}}
\renewcommand{\theequation}{S\arabic{equation}}
\renewcommand{\thepage}{S\arabic{page}}
\setcounter{figure}{0}
\setcounter{table}{0}
\setcounter{equation}{0}
\setcounter{page}{1} % not 0 as \newpage already started a supplementary page
% References continue the numbering from the main text.

%%%%%%%%%%%%%%%% SUPPLEMENT TITLE PAGE %%%%%%%%%%%%%%%

\begin{center}
\section*{Supplementary Materials for\\ \scititle}

% Author list for the supplement
% Indicate the corresponding authors, but do NOT include institutions here
% It would be nice if the template auto-generated this, but doing so is complicated...
	Cody Carr$^{1,2\ast}$,
    Renyue Cen$^{1,2\ast}$,
	Stephan McCandliss$^{3\ast}$,
    Jack Ford$^{3}$,\and
    
    Alberto Saldana-Lopez$^{4}$,
    Claudia Scarlata$^5$,
    Mason S. Huberty$^5$,
    Anne Jaskot$^{6}$,\and
    
    Sophia Flury$^{7}$,
    M. S. Oey$^{8}$,
    Ricardo O. Amor\'{i}n$^{9}$,
    Sanchayeeta Borthakur$^{10}$,\and
    
    Matthew Hayes$^{4}$,
    Timothy Heckman$^{3}$,
    Zhiyuan Ji$^{11}$,
    Lena Komarova$^{8}$,\and
    
    Alexandra Le Reste$^5$,
    Floriane Leclercq$^{12}$,
    Rui Marques-Chaves$^{13}$,
    Leo Michel-Dansac$^{14}$,\and
    
    Göran Östlin$^{4}$,
    Swara Ravindranath$^{15,16}$,
    Michael~J.~Rutkowski$^{17}$,
    Daniel Schaerer$^{10}$,\and
    
    Trinh Thuan$^{18}$,
    Eros Vanzella$^{19}$,
    Bingjie Wang$^{20,21,22}$,
    Xinfeng Xu$^{23}$
    
\small$^\ast$Corresponding author. Email: codycarr24@gmail.com, renyuecen@zju.edu.cn, stephan@pha.jhu.edu\\
%\small$^\dagger$These authors contributed equally to this work.
\end{center}

% Fill out the numbers for each type of supplementary material,
% and delete any lines that aren't applicable.
% These are just example numbers that don't match the rest of this template.
\subsubsection*{This PDF file includes:}
Materials and Methods\\
Supplementary Text\\
References \textit{(69-\arabic{enumiv})}\\
Figures S1-S4\\
Tables S1-S4\\
%Captions for Movies S1 to S2\\
%Captions for Data S1 to S2

%\subsubsection*{Other Supplementary Materials for this manuscript:}
%Movies S1 to S2\\
%Data S1 to S2

\newpage

%%%%%%%%%%%%%%%% MATERIALS AND METHODS %%%%%%%%%%%%%%%

\subsection*{Materials and Methods}

\subsubsection*{Data}

Spectra were obtained with the G130M grating on HST/COS, using CENWAV=1222 for all targets except J103344+635317, which—due to its higher redshift—was observed with CENWAV=1291. G160M data were drawn from previously approved programs (12928, PI Henry; 14201, PI Malhotra; 17153, PI Leclercq) focused on Ly$\alpha$ emission line profiles.  XXM-Newton data was obtained for J103344+635317 from program (0934050101, PI Ji). These spectra, along with photometric data from the LaCOS program (17069, PI Hayes) and Sloan Digital Sky Survey spectra, were corrected for Milky Way foreground reddening with Galactic $E(B-V)$ estimates derived from dust maps \cite{Green2018} and extinction curves \cite{Cardelli1989}. The redshift, orbits, observed flux at 1100 \AA, and the Milky Way extinction are provided for each galaxy in Table~\ref{tab:observations}.

%and-where available-HST/STIS spectra (17169, PI Schaerer), 

\subsubsection*{SED Modeling}

We used the code \textsc{FiCUS}\footnote{\textsc{FiCUS} (Fitting the stellar continuum of Uv Spectra): \url{https://github.com/asalda/FiCUS}} \cite{Chisholm2019, Saldana-Lopez2022, Saldana-Lopez2023} to model the SEDs and measure the light fractions and UV properties provided in Tables~\ref{tab:light_fractions} and \ref{tab:radio_emission_lines}. We followed the procedure outlined by \cite{Saldana-Lopez2022} on the low-resolution G140L data obtained through LzLCS. In \textsc{FiCUS}, the UV stellar continuum modeling is achieved by fitting the observed spectra with a linear combination of theoretical, single-burst \textsc{Starburst99} \cite{Leitherer1999} models. Specifically, we adopted a set of four stellar metallicities (0.05, 0.2, 0.4 and 1.0 $Z_{\odot}$) and ten stellar age bins (0–10 Myr). A nebular continuum was added to every model by self-consistently processing the original \textsc{Starburst99} models through \textsc{Cloudy} v17.04 \cite{Ferland2017}, assuming similar gas-phase and stellar metallicities, an ionization parameter of $\log ~U = -2.5$ and a volume hydrogen density of $n_H = 100~{\rm cm^{-3}}$. We adopted the dust attenuation prescription of \cite{Reddy2016}, and a simple geometry where all the light is attenuated by a uniform foreground slab of dust. 

By combining G130M and G160M spectra, we recovered a wavelength range similar to that probed by G140L, except for a gap of $\sim 50$\,\AA\ that appears in all galaxies except J103344+635317. This wavelength coverage includes key stellar wind features used to constrain stellar ages, such as the O\,VI 1032 \AA, 1038 \AA\ doublet and the N\,V 1238.8 \AA, 1242.8 \AA\ lines.  All recorded quantities (e.g., light fractions, $f_{\rm esc}^{\rm LyC}$, etc.) were obtained directly from the \textsc{FiCUS} outputs, using the same approaches as \cite{Saldana-Lopez2022} and \cite{Flury2022_data}. Importantly, escape fractions ($f_{\rm esc}^{\rm LyC}$) were obtained as the ratio between the ionizing flux measured over the new G130M observations, and the intrinsic flux in the same LyC window predicted by the \textsc{FiCUS} SED fits. 

We measure $f_{\rm esc}^{\rm LyC}$ values that differ from those reported by \cite{Flury2022_data}, with J115855+312559 and J105331+523753 showing the most significant discrepancies. We attribute these differences primarily to the higher spectral resolution of the G130M and G160M data, which allow us to resolve finer details in the spectral lines. Part of the discrepancy may also arise from differences in the masking procedure, specifically the removal of absorption features associated with galactic winds. These features are more easily identified in the higher-resolution spectra. Finally, we note that our $f_{\rm esc}^{\rm LyC}$ measurements generally have smaller uncertainties. These errors were estimated by folding together the uncertainties in the \textsc{FiCUS} SED fits (using 200 iterations) and in the mean LyC flux of our data. Again, the higher resolution of our data returned both higher signal-to-noise in the measured LyC flux and lower uncertainties in the \textsc{FiCUS} best-fit models, which overall reduced the $f_{\rm esc}^{\rm LyC}$ error bars.

We require more data to fully assess how higher spectral resolution data impact our UV SED modeling and measured $f_{\rm esc}^{\rm LyC}$ values. However, here we note that the overall trends involving $f_{\rm esc}^{\rm LyC}$ reported by \cite{Flury2022_data,Flury2022_diagnostics} remain consistent across the LzLCS sample, regardless of the escape fraction measurement method (e.g., the empirical-based flux ratio, $F_{900\text{\AA}}/F_{1100\text{\AA}}$).

%We find that our values for $f_{\rm esc}^{\rm LyC}$ agree with other measurements, including the f900/f1100 ratio and the the escape fraction derived from H$\beta$, $f_{\rm esc}^{\rm LyC}(\rm H\beta)$ for the three strongest leakers in our sample, J103344+635317, J091703+315221 and J115855+312559.  While this should not be expected in general (see \cite{Jaskot2025}) we find it to be rather telling. $f_{\rm esc}^{\rm LyC}(\rm H\beta)$ depends on the global gas distribution in the galaxy as well as the dust content.  These galaxies have low dust extinction in the UV and are expected to have roughly spherical outflow geometries in neutral gas \cite{Carr2025_LyC,Li2025}.  The fact that the three values agree is an indicator that the angular averaged escape fraction and line of sight dependent values are similar.      

\subsubsection*{Spectral Line Fitting}

To estimate the equivalent widths (EWs) and $v_{90}$ values, or the observed velocity at 90\% of the EW when integrating blue-ward of line center, we first fit the spectra using radiative transfer models and then extracted the desired quantities directly from the best-fitting models. This approach reduces the impact of noise and corrects for blue emission infilling in the absorption profiles. We use the semi-analytical line transfer (SALT) radiative transport code \cite{Carr2023} with the Bayesian fitting procedure described in \cite{Carr2021}. SALT has been shown to accurately reproduce a wide range of galactic wind absorption line profiles in both observations \cite{Carr2021,Huberty2024,Carr2025_LyC} and simulations \cite{Carr2023,Carr2025_FIRE2}. A more detailed modeling of these data will be presented in a future paper; additional details on the models can be found in \cite{Carr2023}.

To account for multiple absorption components, we included additional Gaussian-shaped features \cite{Huberty2024}. For example, to model the O VI 1032\AA\ line in J144010 + 461937, we included three components with variable amplitude, width, and central velocity. All measured $v_{90}$ and EW values for the H I 1026 \AA, N III 990 \AA, and O VI 1032 \AA\ lines are listed in Table~\ref{tab:line_data}.  We note that the N III line may suffer from blending with the neighboring O I 988 \AA\ line.  Given the sub-solar metallicities of our sample, and the low oscillator strength for this transition ($\sim 5\times10^{-2}$), we suspect the amount of absorption due to this transition to be small.

We explore correlations using the Kendall rank correlation statistic in Figure~\ref{fig:correlations}. For H I 1026 \AA, we find that $v_{90}$ and the EW show strong and significant correlations with $f_{\rm esc}^{\rm LyC}$ ($\tau = -0.87$, $p = 0.02$). To assess the robustness of these correlations given the measurement uncertainties, we recompute $\tau$ $10^4$ times within uncertainties, using the Monte Carlo approach by \cite{Flury_2023}. We find these correlations to be moderately tight, with a one sigma dispersion in $\tau$ of ($-0.87$--$-0.53$) and ($-0.93$--$-0.6$) around the median. We also find a perfect correlation between the light fraction of stars with ages between $4$--$10$~Myr ($f_*(4 \leq t \leq 10~\rm Myr)$) and $f_{\rm esc}^{\rm LyC}$, although the one sigma dispersion is larger, spanning ($-0.73$--$-0.27$). Similarly, we find a strong correlation between $v_{90}$ and $f_*(4 \leq t \leq 10~\rm Myr)$ ($\tau = -0.87$, $p = 0.02$), with a one sigma dispersion of ($-0.73$--$-0.20$). For this reason, we interpret the latter two correlations as moderate. Together, these relations suggest that $f_{\rm esc}^{\rm LyC}$ declines as a supernova-driven galactic wind develops over $\sim 10$ Myr. On this timescale, continuous mass loading prevents photons from escaping, and longer periods are required for the wind to fully quench star formation and clear the ISM in these galaxies.

Some of these trends are consistent with \cite{Mauerhofer2021}, who examined relationships between mock spectral lines and $f_{\rm esc}^{\rm LyC}$ in simulations. Consistent with our measurements for J144010 + 461937 and J154050 + 572442, they found low $f_{\rm esc}^{\rm LyC}$ ($<2\%$) along sight lines with $v_{90}>700\ {\rm km\ s^{-1}}$ and $\rm{EW}>3$\AA\ in Ly$\beta$. However, they also reported low $f_{\rm esc}^{\rm LyC}$ ($<2\%$) when the Ly$\beta$ absorption profile is dominated by emission infilling, although they noted that the number of lines of sight satisfying this criterion was limited. Our closest analog to this case is J103344+635317, which exhibits a prominent P~Cygni profile yet has $f_{\rm esc}^{\rm LyC} > 20\%$. The lack of absorption in these line profiles suggests an underdeveloped wind. This disagreement in $f_{\rm esc}^{\rm LyC}$ values may arise from the simulation's limited ability to resolve the small-scale physics governing radiation escape in the absence of a SN-driven outflow. We note, however, that the simulation used in their study represents a higher-redshift ($z \sim 3$), lower-SFR galaxy than the galaxies of our sample.

%The corrected spectra were fit for intrinsic dust-attenuated and age weighted SEDs, using  \textsc{FiCUS}\footnote{https://github.com/asalda/FiCUS}  \cite{Chisholm2019, Saldana-Lopez2022, Saldana-Lopez2023}, which incorporates spectral synthesis models from BPASS \cite{Eldridge2017, Stanway2018} and Starburst99 \cite{Leitherer1999} and dust obscuration models for extra galactic sources \cite{Calzetti2000, Reddy2016}. 

%The Materials and Methods section should contain details of the samples measured,
%experiments performed, observations taken, simulations run, data analysis, statistical methods etc.
%Give enough detail for any competent researcher in your field to fully reproduce the results.

%To refer to this section from the main text, use the numbered note in the reference list \cite{methods}.
%Refer to figures and tables in the same way as in the main text but now all capitalized e.g.
%Fig.~\ref{fig:example}, Table~\ref{tab:example},
%Fig.~\ref{fig:sup_example} and Table~\ref{tab:sup_example}.
%Cite references in the usual way \cite{example2},
%including any that are only cited in the supplement \cite{sm_example,conference_example}.

%The numbering of figures, tables, equations and pages has been reset to start from S1, as in
%\begin{equation}
%	\cos(2\theta) = \cos^2\theta - \sin^2\theta.
%	\label{eq:sup_example} % Use a logical label
%\end{equation}

%\subsubsection*{Example supplement heading}

%The two main sections of the supplement can be split up using headings.

%%%%%%%%%%%%%%%% SUPPLEMENTARY TEXT %%%%%%%%%%%%%%%

\subsection*{Supplementary Text}
%The Supplementary Text section can only be used to directly support statements made in the main text
%e.g. to present more detailed justifications of assumptions, investigate alternative scenarios,
%provide extended acknowledgements etc.
%Material in this section cannot claim results or conclusions that weren't mentioned in the main text.
%To refer to this section from the main text, just write (Supplementary Text).

\subsubsection*{The physics of compact, highly star-forming galaxies.}

\begin{figure} % Do NOT use \begin{figure*}
	\centering
	\includegraphics[width=\textwidth]{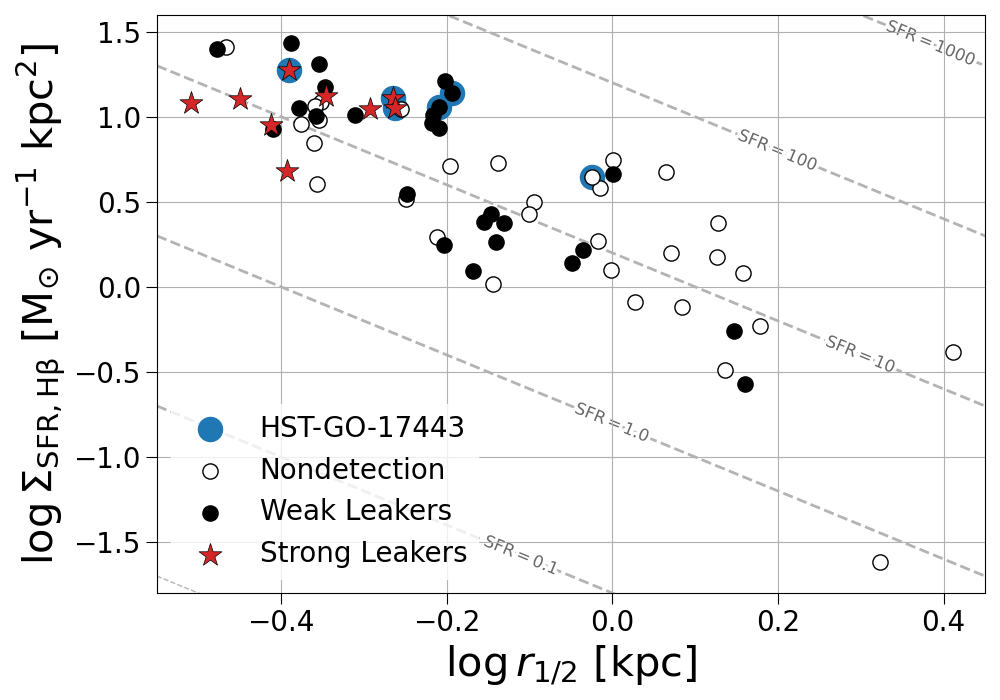} % for an image file named example_figure.*
	% Pick an appropriate width - in print, figures are usually one or two columns wide, which can
	% be approximated by 0.3\textwidth or 0.6\textwidth respectively. Use appropriate label sizes.

	% Captions go below figures
	\caption{\textbf{LzLCS galaxies shown according to star formation surface density ($\Sigma_{\rm SFR}$) and the UV half-light radius ($r_{1/2}$) determined from NUV acquisition imaging in log space.} Strong leakers ($>5\sigma$ LyC detection, $f_{\rm esc}^{\rm LyC}>5\%$), weak leakers ($>2\sigma$ LyC detection, not strong), and non-detections ($<2\sigma$ LyC detection) appear as red stars, black filled circles, and black open circles, respectively.  The galaxies from our sample (HST-GO-17443) appear with a blue shadow.  High $\Sigma_{\rm SFR}$ and compact morphology are associated with strong galactic winds, long proposed to facilitate LyC escape.  However, $f_{\rm esc}^{\rm LyC}$ varies substantially within this region.  Our galaxies were selected from within this parameter space, while controlling for the physical properties that govern the forces driving galactic winds (see Table~\ref{tab:galaxy_props}), to investigate the observed variation in $f_{\rm esc}^{\rm LyC}$. As discussed in the main body of the paper, we attribute this variation to the time evolution of the starburst, with $f_{\rm esc}^{\rm LyC}$ peaking prior to the onset of the outflow. This sequence represents a significant departure from previous theoretical investigations of compact, star-forming galaxies, which found that $f_{\rm esc}^{\rm LyC}$ peaks at later stages—after winds have cleared the galaxy of neutral gas and dust \cite{Kimm2014}. This findings will need to be addressed in the next generation of cosmological simulations of galaxy formation. Data are taken from \cite{Flury2022_data,Flury2022_data_erratum}. See also \cite{Carr2025_LyC}.}
	\label{fig:sigma_r_UV} % give each figure a logical label name
\end{figure}

\begin{figure} % Do NOT use \begin{figure*}
	\centering
	\includegraphics[width=\textwidth]{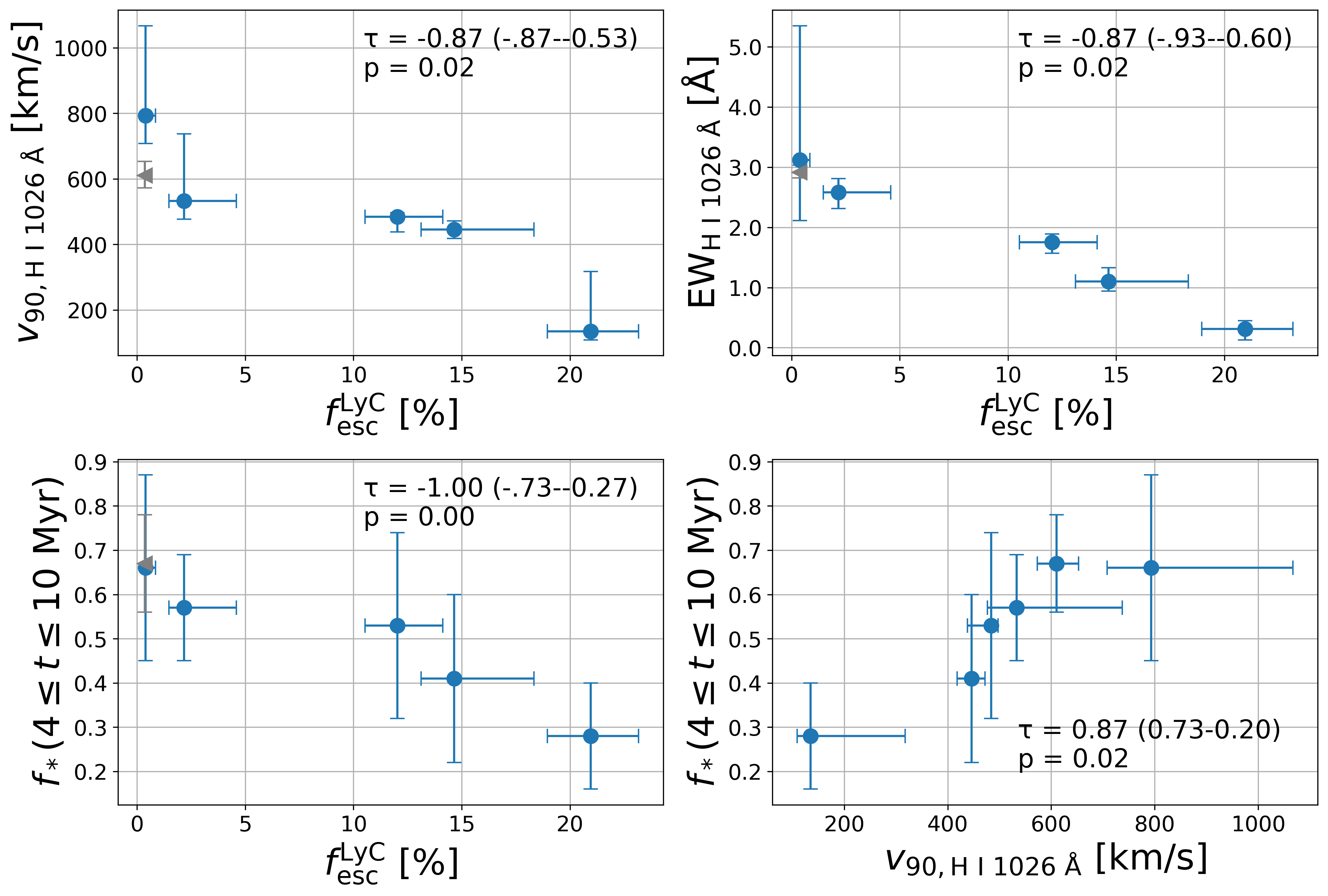} % for an image file named example_figure.*
	% Pick an appropriate width - in print, figures are usually one or two columns wide, which can
	% be approximated by 0.3\textwidth or 0.6\textwidth respectively. Use appropriate label sizes.

	% Captions go below figures
	\caption{\textbf{Kendall Rank Correlations} Upper Left: $v_{90}$ [km/s] versus $f_{\rm esc}^{\rm LyC}$ [\%]; Upper Right: EW versus $f_{\rm esc}^{\rm LyC}$ [\%]; Bottom Left: light fraction $f_*(4\leq t\leq 10~{\rm Myr})$ versus $f_{\rm esc}^{\rm LyC}$ [\%]; Bottom Right: light fraction $f_*(4\leq t\leq 10~{\rm Myr})$ versus $v_{90}$ [km/s].  Upper limits are shown as gray arrowheads. Both $v_{90}$ and EW show strong and significant correlations with $f_{\rm esc}^{\rm LyC}$ ($\tau = -0.87$, $p = 0.02$). Monte Carlo sampling indicates that these correlations are moderately tight, with a one sigma dispersion of ($-0.87$--$-0.53$) and ($-0.93$--$-0.60$)in $\tau$. The light fraction $f_*(4\leq t\leq 10~{\rm Myr})$ shows a perfect correlation with $f_{\rm esc}^{\rm LyC}$ ($\tau = 1.0$, $p = 0.00$). However, the large age uncertainties allow for substantial dispersion, with a one sigma range of ($-0.73$--$-0.27$). A similar result is found for $v_{90}$ versus $f_*(4\leq t\leq 10~{\rm Myr})$ ($\tau = 0.87$, $p = 0.02$), with a one sigma dispersion of ($-0.73$--$-0.20$). For this reason, we conservatively interpret the correlations in the bottom row of panels as moderate.}
	\label{fig:correlations} % give each figure a logical label name
\end{figure}

Our galaxies were selected from LzLCS for their compact sizes and high star formation rates — conditions frequently linked to efficient LyC escape \cite{Cen2020,Naidu2022,Flury2022_data,Carr2025_LyC}. In Figure~\ref{fig:sigma_r_UV}, we compare their star formation rate surface densities ($\Sigma_{\mathrm{SFR}}$) and UV half-light radii ($r_{1/2}$) to the rest of the LzLCS population, labeling each system as a strong leaker ($>5\sigma$ LyC detection, $f_{\rm esc}^{\rm LyC}>5\%$), weak leaker ($>2\sigma$ LyC detection, not strong), or nondetection ($<2\sigma$ LyC detection) according to the LzLCS classification scheme.  Note that these classification schemes, as well as escape fraction values, refer to those measured by \cite{Flury2025_ISM} and differ from our remeasured values and the classification scheme used in Figure~\ref{fig:summary}.  A key objective of our analysis was to understand why galaxies with similar values of $\Sigma_{\mathrm{SFR}}$ and $r_{1/2}$ exhibit such a broad range of $f_{\mathrm{esc}}^{\mathrm{LyC}}$ values.  As we argue, we attribute the differences to the timing and development of the neutral phase of the galactic wind.    

Compact, intensely star-forming galaxies are also known to launch powerful galactic winds, which have long been attributed to their high escape fractions \cite{Heckman2001,Heckman2011,Alexandroff2015,Heckman2016,Cen2020}. These winds are driven by the high pressures associated with clustered supernova explosions and intense radiation fields, which can overcome the inward forces of gravity and the ram pressure of the surrounding medium.  In these galaxies, it is instructive to highlight the important dynamical role of radiation pressure on driving the winds \cite{Cen2020}. We recap the dominant forces involved, illustrated in the spherical case. In reality, the radiation and stellar wind-driven outflows are likely angularly dependent, due to turbulence and subsequently the preferred propagation of radiation and outflows in the directions of least resistance.

%We may calculate a launch condition by balancing forces between a galaxy and an outflow distributed within an extended shell surrounding the galaxy. 
We begin with a spherical, star-forming galaxy with stellar mass $M_{\star}$ and half-light radius $r_{1/2}$. In this context, the outward force produced by supernovae can be expressed as:
\begin{eqnarray}
    F_{\mathrm{SF}} = \mathrm{SFR}\times p_{\mathrm{SN}}\times M_{\mathrm{SN}}^{-1},
\end{eqnarray}
where $p_{\mathrm{SN}} \approx 3\times 10 ^{5}\mathrm{ \ M_{\odot}\ km\ s^{-1}}$ is the terminal momentum generated per supernova \cite{Kimm2014} and $M_{\mathrm{SN}}$ is the amount of mass required to produce one supernova (50,75,100) for the (Chabrier, Kroupa, Salpeter) initial mass functions (IMFs), respectively.  %We ignore the amount of mechanical energy produced in stellar winds, as it is expected to be roughly 10\% of the mechanical energy deposited by supernovae at low metallicities \cite{Jecmen2023}.  
UV plus FIR radiation pressure is important in high star-forming, high-surface density galaxies and will contribute to an outward force on the galaxy which we may compute as 
\begin{eqnarray}
    F_{\mathrm{rad}} = \mathrm{SFR} \times \alpha \times c[1-\exp(-\Sigma_{\mathrm{gas}}\kappa_{\mathrm{UV}})]\times (1+\Sigma_{\mathrm{gas}}\kappa_{\Sigma_{\mathrm{FIR}}}),
\end{eqnarray}
where $\alpha = 3.6\times10^{-4}$ is an adopted nuclear synthesis energy conversion efficiency from rest mass to radiation, $c$ is the speed of light, $\kappa_{\mathrm{UV}} = 1800$ cm${}^2$ g${}^{-1}$ and  $\kappa_{\mathrm{FIR}} = 20$ cm${}^2$ g${}^{-1}$ are the opacity for ultraviolet \cite{Draine2003} and dust-processed FIR radiation \cite{Lenz2017}, respectively, and $\Sigma_{\mathrm{gas}}$ is the surface density of the galaxy.

Countering the above two outward forces are two inward forces: the forces due to gravity and ram pressure.  To simplify our calculations, we assume the wind is isothermal ($\rho \sim r^{-2}$), and compute the mean force of gravity on the surrounding wind to be
\begin{eqnarray}
    F_{\mathrm{g}} = \frac{\ln{(r_{\mathrm{max}}/r_{\mathrm{min}})}GM_{\mathrm{\star}}M_{\mathrm{wind}}(t)}{4r_h^2},
\end{eqnarray}
where $G$ is the gravitational constant, $r_{\mathrm{min}}$ and $r_{\mathrm{max}}$ are the minimum and maximum radii over which the gas is expelled, and $M_{\mathrm{wind}}(t)$ is the mass of the wind.  %We set $r_{\mathrm{min}} = r_{1/2}$ and $r_{\mathrm{max}} = r_{50}^{\mathrm{Ly}\alpha}$, the radius of the Ly$\alpha$ halo at 50\% of its peak surface brightness.  
Finally, we compute the inward force due to the ram pressure of the falling gas in terms of the inflow rate as
\begin{equation}
F_{\mathrm{rp}} = \dot{M_{\mathrm{in}}}v_{\mathrm{in}} = \eta \times \mathrm{SFR} \times \left ( \frac{GM_{\mathrm{gas},0}}{r_{1/2}}\right )^{1/2},
\end{equation}
where $\eta$ is the ratio of the inflow rate to the SFR.  In our sample, we see no evidence of significant inflows of gas \cite{Carr2022}. Thus, the ram-pressure may be unimportant. Since the compact galaxies investigated are in the Super-Eddington regime overall, it is likely that the early intense UV radiation and reprocessed FIR radiation may be able to carve out escape pathways in turbulent gas clouds.

By controlling for SFR, $M_{\star}$, and $r_{1/2}$, we are able to constrain the average forces governing the starbursts in our sample. Controlling for metallicity further helps account for differences in star formation histories and the expected number of SNe per unit stellar mass \cite{Jecmen2023}. The various spectral changes of the winds observed in Figures~\ref{fig:cool_warm_phases} and \ref{fig:stellar_winds} can therefore be attributed primarily to evolutionary effects, occurring on dynamical timescales shorter than that of the SFR [e.g., Figure 8, \cite{Carr2025_LyC}].  Note that given the large covering fractions, we are likely to avoid large fluctuations in $f_{\rm esc}^{\rm LyC}$ due to spatial variations along the line of sight, suggesting our sequence isn't random \cite{Cen2015}.

Given its light-weighted age and large fraction of stars younger than 3 Myr, the galactic wind in J103344 + 635317 is most likely powered by radiation pressure and stellar winds rather than SNe. Some early SNe may still contribute, as the SED fits indicate a modest fraction of older stars ($3$–$10$ Myr). \cite{Martin2024} report bubble expansion at comparable velocities ($\sim$100 km s$^{-1}$) in galaxies with young clusters, which they attribute to early SNe. Yet the absence of detectable X-rays in J103344 + 635317 suggests that, if such events are occurring, they remain embedded within natal birth clouds or are rare, possibly exacerbated by the low metallicities of our sample \cite{Jecmen2023}. Even if a few SNe are present, their energy contribution may be comparable to that of stellar winds, which normally account for only $\sim$10\% of the mechanical energy deposited in the ISM when operating at 100\% conversion efficiency \cite{Leitherer1992,Leitherer1995}.  Thus, while we cannot fully exclude molecular cloud disruption by early SNe, our results confirm that early feedback leads to higher LyC escape, with $f_{\rm esc}^{\rm LyC}$ peaking before the onset of a large-scale galactic wind, capable of suppressing star formation.  These results provide compelling evidence for the growing need to account for early feedback processes in cosmological simulations \cite{Jaskot2017,Jaskot2019,Bait2024,Carr2025_LyC,Flury2025_ISM}.

\subsubsection*{Age Diagnostics}

We provide additional diagnostics for stellar population ages in Table~\ref{tab:radio_emission_lines}. The strongest LyC leaker, J103344 + 635317, exhibits the highest O32 ratio ($[\text{O\,III}]\,\lambda5007 / [\text{O\,II}]\,\lambda\lambda3726,3729 = 0.662$), consistent with a young population dominated by massive O-type stars producing copious ionizing photons and fast stellar winds (up to $\sim 3000~\rm km~s^{-1}$). Its high H$\beta$ equivalent width ($80~\text{\AA}$) and steep UV slope ($\beta_{1550} = -2.6$) indicate youth and low dust obscuration \cite{Chisholm2022}. The flat radio continuum between 3 and 6 GHz (spectral index $\alpha_{3~\rm GHz}^{6~\rm GHz} = 0.13$) suggests free-free emission dominates, with minimal non-thermal synchrotron radiation from SNe \cite{Bait2024}. Follow-up \textit{XMM-Newton} observations resulted in a non-detection, with a 3-sigma upper limit on the 0.5–8 keV X-ray luminosity of $10^{40.8}\ \mathrm{erg\ s^{-1}}$ \cite{Ji2025}. This is below the SFR–metallicity–$L_{\mathrm{X}}$ relation, suggesting that the stellar population is too young for high-mass X-ray binaries to have formed \cite{Brorby2016}. Together, these diagnostics imply that the majority of massive stars are still on the main sequence and that supernova feedback is minimal. In contrast, J154050+572442 exhibits markedly different age diagnostics, with the flattest UV slope ($\beta_{1550} = -1.6$), the lowest $\log(\text{O32})$ ratio ($0.223$), and the lowest $\mathrm{EW}(\mathrm{H}\beta)$ ($40.44~\text{\AA}$) of the sample—all of which are indicative of a more evolved stellar population.  

The next strongest leakers, J091703+315221 and J115855+312559, exhibit mixed age diagnostics, with both very young and more evolved stellar populations. For example, J091703+315221 has the second steepest $\beta_{1550}$ (-2.2), moderate H$\beta$ EW (51 \AA), and a steeper radio spectral index ($-0.23$) than J103344+635317. J115855+312559 shows an even steeper radio slope (-0.91) but still has a prominent $\rm EW(H\beta)=70.22\text{\AA}$.

J105331 + 523753 resembles the intermediate leakers but has an even steeper radio slope ($-0.99$), suggesting a more evolved population, although its large H$\beta$ EW ($73\ \text{\AA}$) indicates this galaxy still hosts a sizable amount of young stars. J144010 + 461937 is similar to J154050 + 572442, with low O32 ($0.38$), modest H$\beta$ EW ($55\ \text{\AA}$), and relatively flat UV slope ($-1.77$), consistent with an older stellar population.

None of these diagnostics are without limitations, and several caveats should be kept in mind. $\beta_{1550\text{\AA}}$ is known to depend strongly on dust—although using $1550\text{\AA}$ instead of $1200\text{\AA}$ should help mitigate this effect \cite{Chisholm2022}. Indeed, the values closely correlate with UV dust extinction in our sample. H$\beta$ is sensitive to stellar mass and is not expected to correlate perfectly with population age; for example, the formation of binary stars can lead to a resurgence in H$\beta$ emission \cite{Flury2022_data}. $\log(\text{O32})$ is sensitive to gas density, metallicity, and even $f_{\rm esc}^{\rm LyC}$ \cite{Jaskot2013,Flury2022_data,Reddy2023}. Finally, the flattening of $\alpha_{3~\rm GHz}^{6~\rm GHz}$ can result from several physical processes, including synchrotron aging, inverse Compton losses, and cosmic ray escape [see \cite{Bait2024} for further explanation]. 

In summary, J103344 + 635317 is likely the youngest galaxy and J154050 + 572442 the oldest.  The remaining galaxies show properties of mixed stellar populations, with J091703 + 315221 and J115855 + 312559 as intermediate young leakers, and J105331 + 523753 and J144010+461937 as the older pair. 

%The age diagnostics: First, UV slope tends to be invariant with age except at the bluest wavelengths (Chisholm+ 2022), so its use as an age diagnostic here is a bit misleading. You mention this at some point in the supplementary text but neither address this caveat nor cite the relevant literature which demonstrates this effect. At 1200 Ang, age and dust are quite degenerate, so I would be reluctant to invoke it as an age indicator. Second, H-beta EW is sensitive to stellar mass in addition to population age and can even be affected by the presence of binaries (e.g., Flury+ 2022a), so it is not always a 1:1 age tracer. Third, O32 is sensitive to other properties, including gas density, metallicity, and even fesc (see Flury+ 2022a,b for discussion on this for LCEs, other references include Jaskot & Oey 2013, Reddy+ 2023). While you’ve largely controlled for the metallicity, the density and fesc are worth noting — high gas density or high fesc can boost O32 without necessarily implying the youngest possible stellar populations. I suggest correcting the text to address these caveats and citing the relevant references.

\subsubsection*{Surface Brightness Profiles}

\begin{figure} % Do NOT use \begin{figure*}
	\centering
	\includegraphics[width=\textwidth]{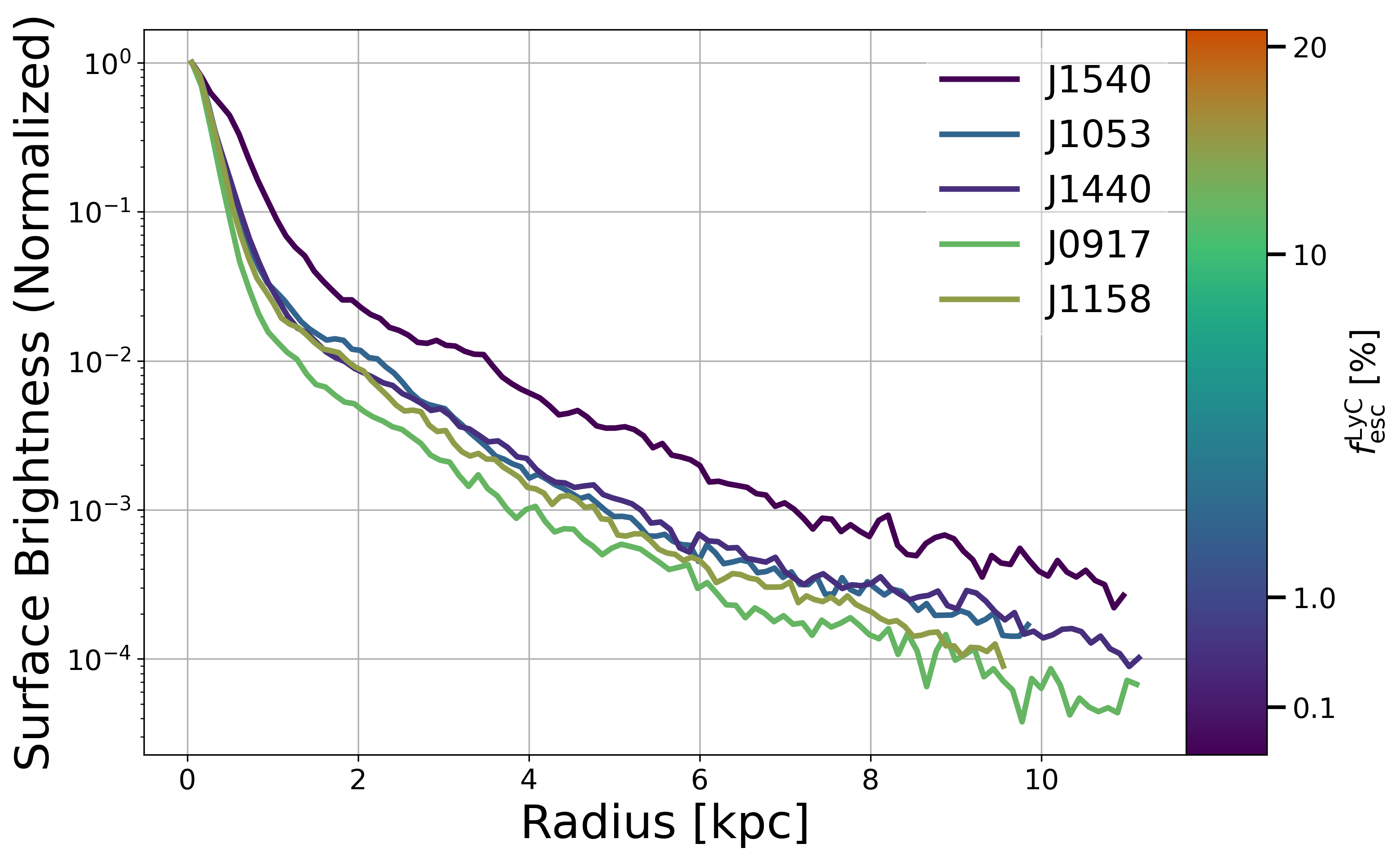} % for an image file named example_figure.*
	% Pick an appropriate width - in print, figures are usually one or two columns wide, which can
	% be approximated by 0.3\textwidth or 0.6\textwidth respectively. Use appropriate label sizes.

	% Captions go below figures
	\caption{\textbf{Normalized surface brightness profiles obtained with the HST/ACS Solar Blind Channel (SBC) using the F150LP filter through LaCOS \cite{Saldana-Lopez2025} for five of the six galaxies in our sample; no data are available for J103344 + 635317.} Galaxies are color-coded by their individual $f_{\rm esc}^{\rm LyC}$ values, as indicated by the color bar. Systems with lower $f_{\rm esc}^{\rm LyC}$ generally exhibit more extended profiles. This trend supports the picture in Figure~\ref{fig:summary}: starbursts in our sample drive fast galactic winds, with the youngest galaxies showing slower outflows that subsequently accelerate and expel gas from the ISM into the CGM in the weakest leakers.}
	\label{fig:SB_profiles} % give each figure a logical label name
\end{figure}

The evolution of the spectral lines presented in Figure~\ref{fig:cool_warm_phases} suggests that the galactic winds in our starbursts are accelerating over time. Traveling several hundred km s$^{-1}$, these winds can transport gas and dust to large distances over 10 Myr. Figure\ref{fig:SB_profiles} shows the normalized surface brightness (SB) profiles of each galaxy (except J103344 + 635317, which has no data), obtained with HST/ACS Solar Blind Channel F150LP via LaCOS\cite{Saldana-Lopez2025}. Our oldest galaxy, J154050 + 572442, is the most extended, consistent with its age and hosting the second-fastest wind in our sample. The next most extended galaxies, J105331 + 523753 and J144010 + 461937, correspond to the next oldest pair, while J091703 + 315221 and J115855 + 312559 have the shallowest profiles, in agreement with their slower outflows and younger ages. This correlation between age, spatial extent, and wind speed also corresponds to a decline in $f_{\rm esc}^{\rm LyC}$, consistent with the observed trend that $f_{\rm esc}^{\rm LyC}$ decreases with Ly$\alpha$ halo extent \cite{Saldana-Lopez2025}. Our general picture suggests that $f_{\rm esc}^{\rm LyC}$ wanes over $\sim 10$ Myr as galaxies expel gas and dust from the ISM into the CGM.  While these results support our conclusions, we note that the SB profiles likely detect gas outside the stellar winds, including accretion from the intergalactic medium and galaxy-to-galaxy transfers.  

\subsubsection*{Comparison to Literature}

\cite{Amorin2024,Li2025,Komarova2025} report positive correlations between $f_{\rm esc}^{\rm LyC}$ and wind velocity (or related quantities), while we find the opposite.  Here, we examine these studies in more detail relative to our results.  

\begin{figure} % Do NOT use \begin{figure*}
	\centering
	\includegraphics[width=\textwidth]{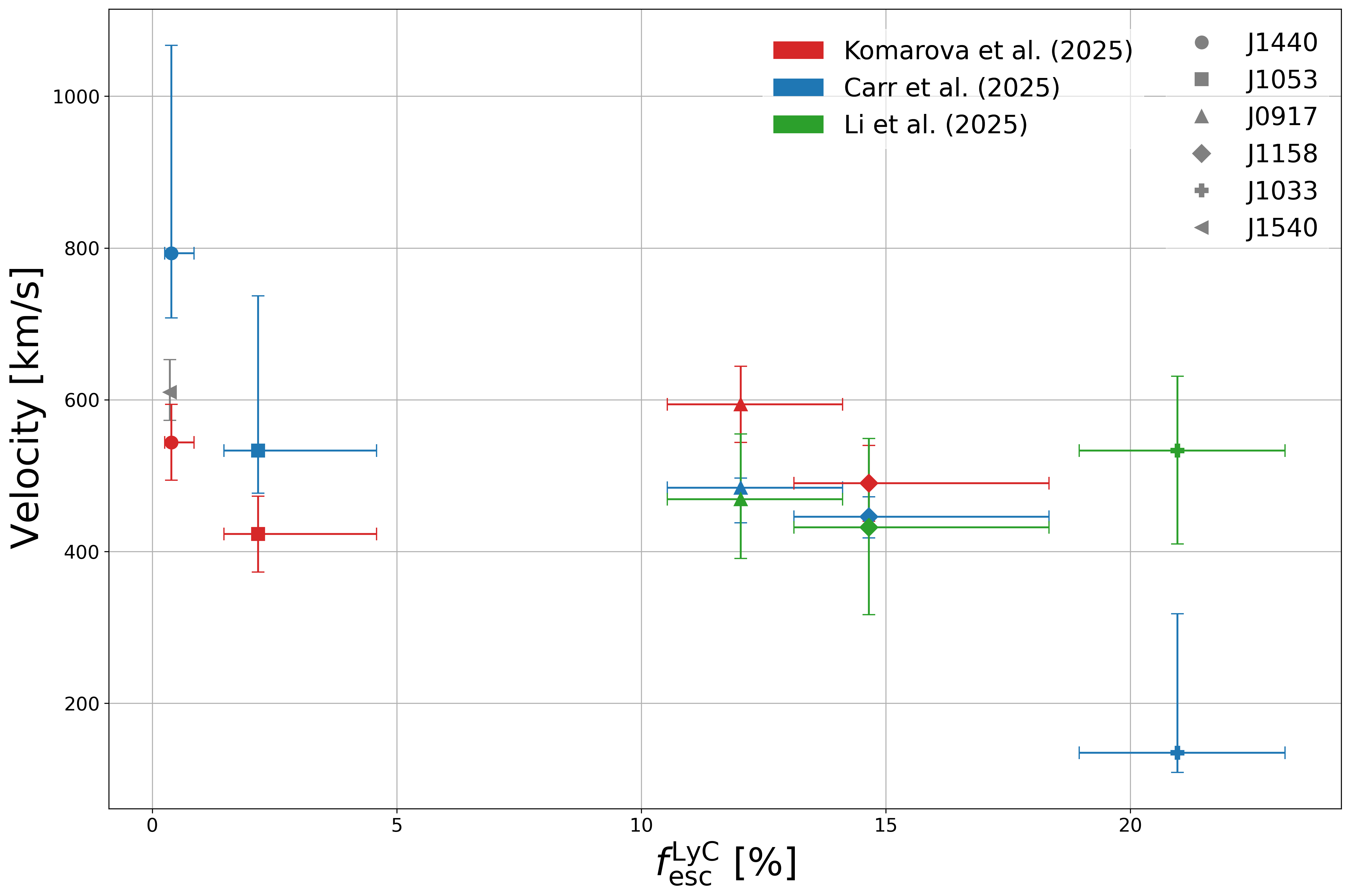}
	\caption{\textbf{Comparing velocity versus $f_{\rm esc}^{\rm LyC}$ using different velocity estimates from the literature.} H I 1026 \AA\ $v_{90}$ values from this work are shown in light blue. $V_{\rm max}$ values derived from the [O III] 5007 \AA\ line by \cite{Komarova2025} are shown in red, while $V_{\rm max}$ values derived from the Mg II 2800 \AA\ doublet by \cite{Li2025} appear in green. The lone grey point corresponds to J154050+572442 and represents an upper limit in $f_{\rm esc}^{\rm LyC}$; data for this galaxy are available only in this work. The \cite{Komarova2025} estimates generally agree with our measurements within the error bars, except for J144010+461937 and J105331+523753, where their values underestimate ours. We attribute this to cooling in older starbursts, which produces lower-density gas that is less easily detected in emission lines. \cite{Li2025} measurements also generally agree with ours, except for J103344+635317, where they overestimate our value. We attribute this to the low spectral resolution of their study and the difficulty of estimating maximum velocities from line profiles lacking strong absorption components, as is the case for J103344+635317.}
	\label{fig:vel_comp} % give each figure a logical label name
\end{figure}

Both \cite{Amorin2024,Komarova2025} find a positive correlation between $f_{\rm esc}^{\rm LyC}$ and [O III] 5007 \AA\ broad line wings, with \cite{Komarova2025} defining a maximum velocity ($V_{\rm max}$) where the broad component of the emission wing intercepts the continuum.  We compare their values to our H I 1026 \AA\ $v_{90}$ measurements in Figure~\ref{fig:vel_comp} for galaxies with overlapping data, J115855+312559, J091713+315221, J105331+523753, J144010+461937.  We find that our values agree within error bars except for the cases of J144010 + 461937 and J105331 + 523753, where \cite{Komarova2025} measure smaller values.  We also note that the classification of winds as radiation-driven, SN-driven, or ambiguous by \cite{Komarova2025}, based on the morphology of the [O III] 5007 Å broad wings, is consistent with our findings. In fact, all four galaxies in our sample with overlapping coverage are classified by them as either SN-driven or ambiguous.    

These results are consistent with the trends reported by \cite{Xu2025} who found that the wind speeds of emission lines (e.g., H$\alpha$) and absorption lines (e.g., Si II, C II) correlate, with emission lines typically reporting smaller observed velocities than absorption lines in the highest-velocity cases.  \cite{Xu2025} also investigate the gas distribution recovered from each line, finding smaller spatial extents recovered from emission lines.  They attribute these results to a radial decline in density and a
corresponding increase in outflow velocity (i.e., accelerating winds), combined with the fact that emission line luminosity scales with the square of the density while absorption line depth scales only linearly.  While this explanation likely holds in a broad sense for our case as well, we note that our O VI and N III absorption lines show evidence of multiple components in the weakest leakers of our sample, J144010+461937 and J154050+572442, suggesting that higher ionization state emission lines may be sensitive to different regions of the outflows.

We also plot the maximum velocity estimates for overlapping galaxies studied by \cite{Li2025} in Figure~\ref{fig:vel_comp}. These values are model-dependent and derived from radiative transfer modeling of the Mg II 2800 \AA\ doublet. In general, they agree with our Ly$\beta$ $v_{90}$ measurements, except for J103344+635317, where \cite{Li2025} report a significantly faster velocity ($V_{\rm max} = 533\ \rm km/s$). Their study was performed at lower resolution ($\sim 90\ \rm km/s$), where the Mg II absorption lines were not detectable in absorption \cite{Carr2025_LyC}, leaving only an emission profile as the primary component of the line profile to fit with their model (see their Figure 18). Under these conditions, it is difficult to accurately determine maximum wind velocities.  See \cite{Carr2025_LyC} who discuss a similar scenario where the maximum velocity of a wind can go unconstrained from above if the density field decays to an undetectable level in absorption before it is reached. Given that weak absorption features are expected in the spectra of strong LyC emitters, this scenario may explain why they find the opposite correlation from us.  

Ultimately, higher-resolution data are required to fully reconcile the differences between the trends reported by \cite{Amorin2024,Li2025,Komarova2025} and this work. This is especially true for strong LyC leakers, where such data are essential for detecting their relatively weak absorption features in neutral hydrogen lines. Lastly, we emphasize the importance of the parameter control used to isolate the LyC escape sequence observed in this study. For example, wind velocities can vary substantially with stellar mass, SFR, and other galaxy properties, which could obscure the intrinsic relationship between outflows and $f_{\rm esc}^{\rm LyC}$.

%High-resolution [O III] data are unavailable for J103344 + 6353417 and J154050 + 572442, but for the remaining galaxies, the velocities measured by \cite{Komarova2025} generally agree with our $v_{90}$ values within uncertainties, except for J144010 + 461937 ($v_{\rm [O\ III]\ 5007\text{\AA}} = 550\pm 50\ \rm km\ s^{-1} < v_{90,\rm H\ I\ 1026\ \text{\AA}}=793^{+274}_{-85}$). Slower [O III] emission-line velocities compared to H I absorption may indicate cooling in the high-velocity gas near the galaxy outskirts, consistent with our observations of J154050 + 572442. Moreover, the radiation-driven winds inferred by \cite{Komarova2025} are expected to be very diffuse \cite{Komarova2021}, leaving open the possibility of fast, early radiation-driven winds in J103344 + 6353417 that remain undetected in absorption. This may explain why we measure $f_{\rm esc}^{\rm LyC}$ to decline with increasing H I wind velocity, whereas \cite{Amorin2024,Komarova2025} find the opposite trend with [O III].

%Komarova: j0917--610, J1440--518, J1158 519, J1053--550, 

% If your supplement is very short you might need to uncomment the following line to avoid
% layout problems with the figures and tables.
\newpage

%%%%%%%%%%%%%%%% SUPPLEMENTARY FIGURES %%%%%%%%%%%%%%%

%%%%%%%%%%%%%%%% SUPPLEMENTARY TABLES %%%%%%%%%%%%%%%

\begin{table}
    \centering
    \caption{\textbf{Target properties for the LzLCS galaxies.}
    Redshifts ($z$), FUV fluxes at 1100\AA, Milky Way dust extinction, number of HST orbits required for G130M spectroscopy, and mean background subtracted LyC flux measured between 880-900 \AA, except for J103344+635317 which was measured in a 20\AA\ window below 1180\AA\ (observed frame) to avoid geocoronal contamination.  Upper limits represent one sigma uncertainties.  Fluxes are in units of 10$^{-17}\rm erg\ s^{-1}\ cm^{-2}$ \AA${}^{-1}$. }
    \label{tab:observations}
    
    \begin{tabular}{lccccc}
        \hline
        Object & $z$ & $F_{1100}$ & E(B$-$V)$_{\rm MW}$ & Orbits&$F_{\text{LyC}}$ \\
               &       & ($10^{-17}$)         & (mag)             &        & ($10^{-17}$) \\
        \hline
        J103344+635317 & 0.347 & 50.1 & 0.016 & 5 & $10.24^{0.50}_{0.22}$\\
        J115855+312559 & 0.243 & 84.8  & 0.023 & 3 & $23.55^{5.70}_{1.86}$\\   
        J091703+315221 & 0.300 & 43.1 & 0.020 & 5 & $6.20^{0.80}_{0.29}$\\
        J105331+523753 & 0.253 & 122.0 & 0.019 & 2 & $3.96^{4.41}_{1.24}$\\
        J144010+461937 & 0.301 & 70.5 & 0.018 & 4 & $0.77^{0.90}_{0.25}$\\
        J154050+572442 & 0.294 & 72.3  & 0.019 & 4 & $<0.78$\\
        \hline
    \end{tabular}
\end{table}

\begin{table}
    \centering
    \caption{\textbf{Stellar Age Diagnostics}
    Columns show the radio spectral index between 3 GHz and 6 GHz, the equivalent width of H$\beta$, the ionization-sensitive O32 line ratio, and UV spectral index near 1550 \AA.  %, and the Ly$\alpha$ halo half-light radius.  
    Rows correspond to the same galaxy sample as in Tables~\ref{tab:galaxy_props} and \ref{tab:light_fractions}.  Radio information was obtained from \cite{Bait2024} and the remaining information was obtained from \cite{Flury2022_data,Flury2022_data_erratum}.}
    \label{tab:radio_emission_lines}
    
    \begin{tabular}{lcccccc}
        \hline
        Object &$\mathrm{E(B-V)_{UV}}$& $\alpha^{\mathrm{3GHz}}_{6\,\mathrm{GHz}}$ & $\mathrm{EW}(\mathrm{H}\beta)$ & $\log(\mathrm{O32})$ & $\beta_{1550\,\text{\AA}}$ \\%& $R_{50}^{\rm Ly\alpha}$\\
               &         &                      & (\AA)             & (dex)             & & \\%(kpc) \\
        \hline
        J103344+635317 &$0.027\pm0.011$&$0.129 \pm 0.019$  & $80.07 \pm 1.505$ & $0.662 \pm 0.030$ & $-2.642 \pm 0.057$ \\%& --  \\
        J115855+312559 &$0.118\pm 0.007$&$-0.907 \pm 0.108$ & $70.22 \pm 1.451$ & $0.374 \pm 0.023$ & $-1.957 \pm 0.036$ \\%& $1.08 \pm 0.07$\\
        J091703+315221 &$0.068\pm 0.013$&$-0.227 \pm 0.046$ & $50.68 \pm 0.798$ & $0.422 \pm 0.023$ & $-2.167 \pm 0.067$ \\%& $2.26 \pm 0.33$\\
        J105331+523753 &$0.130\pm 0.006$&$-0.992 \pm 0.111$ & $73.36 \pm 0.863$ & $0.531 \pm 0.023$ & $-1.893 \pm 0.030$ \\%& $4.67 \pm 0.42$\\
        J144010+461937 &$0.161\pm0.013$&$-0.717 \pm 0.076$ & $54.65 \pm 0.698$ & $0.382 \pm 0.023$ & $-1.767\pm 0.064$ \\%& $2.17 \pm 0.22$\\
        J154050+572442 &$0.194\pm 0.011$&--                 & $40.44 \pm 0.777$ & $0.223 \pm 0.029$& $-1.593\pm 0.062$\\ %& $4.27 \pm 0.61$ \\
        \hline
    \end{tabular}
\end{table}

\begin{table}
    \centering
    \caption{\textbf{Velocity and Equivalent Width (EW) metrics.} $v_{90}$ values correspond to the observed velocity measured at 90\% the EW when integrating blue-ward of line center.  All values were determined with the SALT model, using pure absorption spectra.}
    \label{tab:line_data}
\begin{tabular}{lcc|cc|cc}
\hline\hline
Galaxy &
EW$_{\rm 1026}$ & $v_{90,{\rm 1026}}$ &
EW$_{\rm 990}$ & $v_{90,{\rm 990}}$ &
EW$_{\rm 1032}$ & $v_{90,{\rm 1032}}$ \\
\hline
& (\AA) & (km s$^{-1}$) & (\AA) & (km s$^{-1}$) & (\AA) & (km s$^{-1}$)\\
J103344+635317 & $0.31^{+0.14}_{-0.18}$ & $135^{+183}_{-26}$ & $0.13^{+0.20}_{-0.03}$ & $194^{+169}_{-19}$ & $0.37^{+0.10}_{-0.17}$ & $176^{+63}_{-10}$ \\
J115855+312559 & $1.10^{+0.23}_{-0.16}$ & $446^{+26}_{-28}$ & $1.05^{+0.03}_{-0.17}$ & $569^{+31}_{-26}$ & $0.43^{+0.12}_{-0.17}$ & $410^{+90}_{-153}$ \\
J091703+315221 & $1.75^{+0.14}_{-0.18}$ & $484^{+13}_{-46}$ & $0.85^{+0.16}_{-0.19}$ & $410^{+13}_{-45}$ & $0.96^{+0.05}_{-0.17}$ & $471^{+14}_{-92}$ \\
J105331+523753 & $2.58^{+0.23}_{-0.27}$ & $533^{+204}_{-56}$ & $1.23^{+0.03}_{-0.25}$ & $634^{+192}_{-77}$ & $0.58^{+0.01}_{-0.13}$ & $483^{+7}_{-170}$ \\
J144010+461937 & $3.12^{+0.23}_{-1.01}$ & $793^{+274}_{-85}$ & $1.70^{+2.24}_{-0.01}$ & $837^{+270}_{-86}$ & $1.17^{+0.06}_{-0.02}$ & $733^{+13}_{-8}$ \\
J154050+572442 & $2.91^{+0.12}_{-0.09}$ & $610^{+43}_{-37}$ & $2.08^{+0.11}_{-0.08}$ & $658^{+43}_{-40}$ & $1.06^{+0.23}_{-0.38}$ & $424^{+24}_{-12}$ \\
\hline
\end{tabular}
\end{table}

%\begin{table}
%    \centering
%    \caption{\textbf{Ly$\alpha$ halo spatial extent and age constraints.}
%    Ly$\alpha$ halo radius at 50\% in surface brightness are shown in units of kpc and ages in Myr for inflow rate to SFR ratios of $\eta = 0,10,\ \text{and}\ 100$.}
%    \label{tab:ages}
    
%    \begin{tabular}{lcc}
%        \hline
%        Object & $R_{50}^{\rm Ly\alpha}$ & age\\
%               & (kpc) & (Myr)\\
%        \hline
%        J103344+635317 & --  & -- \\
%        J091703+315221 & $2.26 \pm 0.33$ &\\
%        J115855+312559 & $1.08 \pm 0.07$& \\
%        J105331+523753 & $4.67 \pm 0.42$& \\
%        J144010+461937 & $2.17 \pm 0.22$& \\
%        J154050+572442 & $4.27 \pm 0.61$&\\
%        \hline
%    \end{tabular}
%\end{table}

%\begin{table} % Do not use \begin{table*}
%	\centering
	% Captions go above tables
%	\caption{\textbf{All captions must start with a short bold sentence, acting as a title.}
%		Follow the same style as main text tables.
%		If the design is similar to previous tables, avoid repetition by refering back to them.}
%\label{tab:sup_example} % give each table a logical label name
%\end{table}

%	\begin{tabular}{lccr} % four columns, alignment for each
%		\\
%		\hline
%		A & B & C & D\\
%		\hline
%		1 & 2 & 3 & 4\\
%		2 & 4 & 6 & 8\\
%		3 & 5 & 7 & 9\\
%		\hline
%	\end{tabular}
%\end{table}     

%%%%%%%%%%% CAPTIONS FOR OTHER SUPPLEMENTARY FILES %%%%%%%%%%

\clearpage % Clear all remaining figures and tables then start a new page

%\paragraph{Caption for Movie S1.}
%\textbf{All captions must start with a short bold sentence, acting as a title.}
%Then explain what is shown in the supplementary video file.
%Give as much detail as you would for a figure e.g. explain axes, color maps etc.
%If the video is an animated equivalent of one of the static figures, state e.g.
%`Animated version of Figure~\ref{fig:example}.'

%\paragraph{Caption for Data S1.}
%\textbf{All captions must start with a short bold sentence, acting as a title.}
%Then explain what is included in the supplementary data file.
%Give as much detail as you would for a table e.g. explain the meaning of every column,
%units used, any special notation etc.

%%%%%%%%%%%%%%%% SUPPLEMENTARY REFERENCES %%%%%%%%%%%%%%%

% Do NOT include a reference list in the supplement.
% All references must be in a single list at the end of the main text.
% The copyeditors will ensure that the correct reference list appears with each version of the paper
% (print, HTML, PDF, mobile app, metadata for bibliographic databases etc.)

\end{document}